\documentclass[aps,pra,twocolumn,showpacs,amsmath,amssymb,floatfix,superscriptaddress,notitlepage]{revtex4-1}

\usepackage{color}
\usepackage{bbm}
\usepackage{graphicx}% Include figure files
\usepackage{dcolumn}% Align table columns on decimal point
\usepackage[utf8]{inputenc}
\usepackage{float}

\usepackage[shortlabels]{enumitem}
\usepackage{graphicx}
\usepackage{epstopdf}
\usepackage{amsthm}
\usepackage{amsmath}
\usepackage{empheq}
\usepackage{bbm}
\usepackage{braket}
\usepackage{amssymb}
\usepackage[usenames,dvipsnames]{xcolor}
\usepackage{cases}
\usepackage{latexsym}
\usepackage[colorlinks=true,citecolor=Cerulean,linkcolor=RubineRed,urlcolor=Cerulean]{hyperref}
\usepackage{subfigure}

\graphicspath{ {images/}{images/figure1/}{images/figure2/}{images/figure3/}{images/figure4/}{images/figure5/} }

\newtheorem{proposition}{Proposition}
\newtheorem{lemma}{Lemma}

\newtheorem{result}{Result}
\newtheorem{definition}{Definition}

\newcommand{\tr}[1]{\operatorname{\textnormal{Tr}}\left[ {#1} \right]}  %Trace
\usepackage{lipsum}
\usepackage{graphicx}

\newcommand{\suppress}[1]{}

\newcommand{\e}{\mathrm{e}}

\newcommand{\dist}{\mathrm{dist}}

\newcommand{\norm}[1]{\left\Vert #1 \right\Vert}

\begin{document}

\title{Locally accurate tensor networks for thermal states and time evolution}
\date{\today}

\author{\'Alvaro M. Alhambra}
\email{alvaro.alhambra@mpq.mpg.de}
\affiliation{Max-Planck-Institut für Quantenoptik, Hans-Kopfermann-Straße 1, D-85748 Garching, Germany}
\affiliation{Munich Center for Quantum Science and Technology (MCQST), Schellingstr. 4, D-80799 München, Germany}

\author{J. Ignacio Cirac}
\affiliation{Max-Planck-Institut für Quantenoptik, Hans-Kopfermann-Straße 1, D-85748 Garching, Germany}
\affiliation{Munich Center for Quantum Science and Technology (MCQST), Schellingstr. 4, D-80799 München, Germany}

\begin{abstract}
Tensor network methods are routinely used in approximating various equilibrium and non-equilibrium scenarios, with the algorithms requiring a small bond dimension at low enough time or inverse temperature. These approaches so far lacked a rigorous mathematical justification, since existing approximations to thermal states and time evolution demand a bond dimension growing with system size. To address this problem, we construct PEPOs that approximate, for all local observables, \emph{i)} their thermal expectation values and \emph{ii)} their Heisenberg time evolution. The bond dimension required does not depend on system size, but only on the temperature or time. We also show how these can be used to approximate thermal correlation functions and expectation values in quantum quenches. 
\end{abstract}
\maketitle

\section{Introduction}

The classical simulation of quantum many-body systems is an important challenge for many different fields, including condensed matter physics, quantum chemistry, quantum information and high energy physics. Approximating generic settings efficiently is widely believed to be impossible, due to the exponential growth of the Hilbert space dimension with the system size. However, many situations of interest do not occur on generic regions of the Hilbert space, but are rather confined to the ``physical corner" of it . This can then be covered by appropriate variational ans\"atze, with tensor networks being the most prominent example. 

Indeed, tensor network methods based on the DMRG algorithm \cite{White92} are routinely used for the simulation of many important physical situations. 
Most prominently, they are used for low energy properties in one and even two dimensions, with great success \cite{Schollw_ck_2011}. They are also widely used in the approximation of finite temperature phenomena \cite{Verstraete_2004,Or_s_2008,White_2009,Stoudenmire_2010,Li_2011,Binder_2015,Czarnik_2015,Chen_2017,Chen_2018,Kshetrimayum_2019,chung2019minimally}, and in the simulation of dynamics \cite{Daley_2004,Feiguin_2005,Manmana_2005,Garc_a_Ripoll_2006,Vidal_2007,Haegeman_2011,Wall_2012,Karrasch_2014,Binder_2015,Zaletel_2015,Haegeman_2016,Ronca_2017,Paeckel_2019,Vanhecke_2021} for short times. This allows for the computation of properties on large system sizes in many situations of interest.

These methods are supported by a series of mathematically rigorous results. For low energies, it is known that ground states of gapped models in 1d have good matrix product state (MPS) approximations \cite{Hastings_2007Area,arad2013area,huang2015area}, and that these approximations can be found efficiently with explicit algorithms \cite{Landau2015,huang2015polynomialtime,Chubb_2016} (see \cite{Beaudrap_2010,anshu2021area} for current progress in two dimensions). For thermal states $\propto e^{-\beta H}$, it is known that they can be approximated in any dimension by tensor networks if $\beta$ is not too large  \cite{Hastings_2006,Molnar_2015,Kliesch_2014,kuwahara2020improved}. Similar results are also known for the unitary time evolution at short times $e^{-itH}$ \cite{Osborne_2006,Hastings_2008,kuwahara2020improved}. All of these previous works aim at approximating the \emph{whole} ground state, thermal state or unitary, respectively. This can be achieved with a bond dimension that grows with system size.

 However, for many physical applications, such as calculating local order parameters, one does not necessarily require a full global approximation, but just a tensor network that describes the relevant local properties well. The success of existing numerical implementations suggests that a much smaller bond dimension, independent of system size, is required in this case.
 
This problem has been previously explored for ground states: that such local approximations exist in 1d gapped models has been shown in a mathematically rigorous way. First, with matrix product operators (MPOs) \cite{huang2015computing,schuch2017matrix} and more recently with MPS \cite{Dalzell_2019,huang2019approximating,huang2020computing}, as well as with projected entangled pairs (PEPS) for 2D ground states with an area law \cite{huang2019approximating,huang2020computing} (see \cite{Huang2019matrixproductstate} for a perspective). For thermal states and time evolution, previous results indicate that it is possible to simulate specific local properties in an efficient way \cite{Hastings_2008,Kliesch_2014}. However, it was previously not known whether there exist particular tensor networks that approximate \emph{all} the local properties of a system with a bond dimension independent of system size.

Here we address this question, by constructing tensor networks with a provably small bond dimension that approximate, for \emph{any} local operator $A$ and in any spatial dimension:
\begin{itemize}
    \item Thermal expectation values $\langle A \rangle_\beta \equiv\tr{A \frac{e^{-\beta H}}{Z}}$.
    \item The Heisenberg time evolution $e^{-itH}Ae^{itH}$.
\end{itemize}
  By linearity, they also approximate extensive sums of local observables $A=\frac{1}{N} \sum^N_{x=0} A_x$. The results hold for Hamiltonians $H$ that are short-ranged, but not necessarily translation-invariant. The bond dimension has a similar dependence on $\beta$ and $t$ as previous global approximations, but it now does not grow with system size. 
  
  Notably, our constructions are explicit, and give rise to algorithms that can in principle be implemented in practice. While these are likely less efficient or more cumbersome to implement than other known methods used in practice, the advantage is that we have performance guarantees. These do not currently exist for most algorithms used in practice, such as the paradigmatic DMRG algorithm. Theoretical guarantees for certain algorithms support the fact that state-of-the-art methods give accurate results. This is because said guarantees show that the methods target quantities that can in principle be computed efficiently.
 
 We prove these guarantees with the aid of previous results on global approximations \cite{Osborne_2006,Hastings_2006,Kliesch_2014,Molnar_2015,kuwahara2020improved}, combined with ideas that allow us to exploit the locality of the problem. For thermal states, this is the principle usually known as the \emph{local indistinguishability} \cite{Michalakis_2013,Kliesch_2014,Schwarz_2017,brandao2019finite}, which relies on the \emph{clustering of correlations} \cite{brandao2019finite,bluhm2021exponential}. For time evolution it is the Lieb-Robinson bound \cite{lieb1972finite,Bravyi_2006}. We also introduce a tensor network construction of a linear map that outputs different PEPO approximations for the unitary dynamics of local operators depending on their support, which may be of independent interest.

We then show how these results allow us to compute quantities of interest.  We focus on approximations to auto-correlation functions (such as current operators in transport problems \cite{Barthel_2009,Karrasch_2012,Karrasch_2013,Barthel_2013,Tiegel_2014,Karrasch_2015,bertini2020finitetemperature}), and time-dependent local expectation values in quantum quenches \cite{Karrasch_2014,White_2018,leviatan2017quantum,Kloss_2018,Paeckel_2019}. These two are particularly relevant to current experiments in quantum simulation platforms such as cold atoms, superconducting qubits or trapped ions, since they are some of the most easily measurable and informative quantities.

The paper is structured as follows. First, we explain the definitions and the setting in Sec. \ref{sec:setting}. Then, we show our result for local thermal states in Sec. \ref{sec:thermal}, and for time evolution in Sec. \ref{sec:timeevol}. We explain the impact of our results for correlation functions and quantum quenches in \ref{sec:appl}, and conclude. The technical proofs and further background are placed in the Appendices.

%%%%%%%%%%%%%%%%
%%%%%%%%%%%%%%%%

\section{Setting and definitions}\label{sec:setting}

Throughout this work, the notions of approximation used are in terms of closeness in $1$-norm or trace norm $\norm{X}_1$ for quantum states and their PEPO approximations \cite{Gilchrist_2005}, and the operator norm $\norm{X}$ for operators. The big-$\mathcal{O}$ notation indicates that a quantity $f=\mathcal{O}(n)$ is such that for some constant $c$, $f \le c n$. $\tilde{\mathcal{O}}(n)$ indicates polylogarithmic corrections $f \le c n \times \text{polylog}(n)$, and $o(n)$ that the scaling is strictly smaller than linear in $n$.

We focus on systems governed by a local Hamiltonian $H=\sum_x h_x$ with uniformly bounded, short range interactions $\max_x \norm{h_x} \le h$ between $N$ particles of small local dimension. The interactions have the graph structure of a $d$-dimensional lattice $\Lambda$ with growth constant $\gamma$ and maximum degree $z$. We denote the small connected regions we focus on as  $\mathcal{R}$, which have a maximum length $k$, such that $\vert \mathcal{R} \vert \le k^d$ (that is, the small region can be embedded on a hypercube of length $k$).

The operators that approximate $e^{-itH}$ and $e^{-\beta H}$ are MPOs and their higher-dimensional generalization PEPOs \cite{Pirvu_2010} which are operator generalizations of MPS and PEPS, respectively. In $d=1$, an MPO $M_D$ of bond dimension $D$ can be written simply as \cite{Verstraete_2004,Zwolak_2004,Pirvu_2010}
\begin{align}
M_D = \sum^N_{\substack{s_1,s_2,\ldots,s_n =1 \\ s_1',s_2',\ldots,s_n' =1}}& \tr{ B_{1}^{[s_1,s_1']}B_{2}^{[s_2,s_2']} \cdots B_{n}^{[s_n,s_n']} } \notag \\
&\ket{s_1,s_2,\ldots,s_n}\bra{s_1',s_2',\ldots,s_n'} , 
\label{Matrix_Product_Op_Def}
\end{align}
where each of the matrices $\{B_i^{[s_i,s_i']}\}_{i,s_i,s_i'}$ is of dimension $D\times D$.

On the other hand, PEPOs are defined in terms of the interaction graph with edges $\{ e \} \in \mathcal{E}$ and vertices $\{v \} \in \mathcal{V}$ as \cite{Verstraete2004,Pirvu_2010,Molnar_2015}
\begin{equation}\label{eq:PEPO}
    M_D = \sum_{\alpha: \mathcal{E} \rightarrow \{1,...,D\}} \bigotimes_{v \in \mathcal{V}} X^v_{\alpha(e^{v}_1),...,\alpha(e^{v}_z(v))},
\end{equation}
where $X^v_{\alpha(e^{v}_1),...,\alpha(e^{v}_z(v))}$ is an operator acting on vertex $v$, $z(v)$ is its degree and $e^v_1,..., e^{v}_z(v)$ are the vertices going through it. %That is, for each vertex $v$ with $z(v)$ edges, there are at most $D^{z(v)}$ distinct operators in the sum Eq. \eqref{eq:PEPO} acting on it.
See \cite{Pirvu_2010,Molnar_2015} for more detailed descriptions. 

We also introduce the notion of a \emph{PEPO map}, which appears in one of our main results (Result \ref{th:timehighd}). This is a linear map $\mathcal{M}(A)$ that takes a PEPO $A$ as an input, and outputs another PEPO with a potentially larger bond dimension. Schematically, it can be understood as 

\begin{figure}[H]
\centering
\includegraphics[width=1\linewidth]{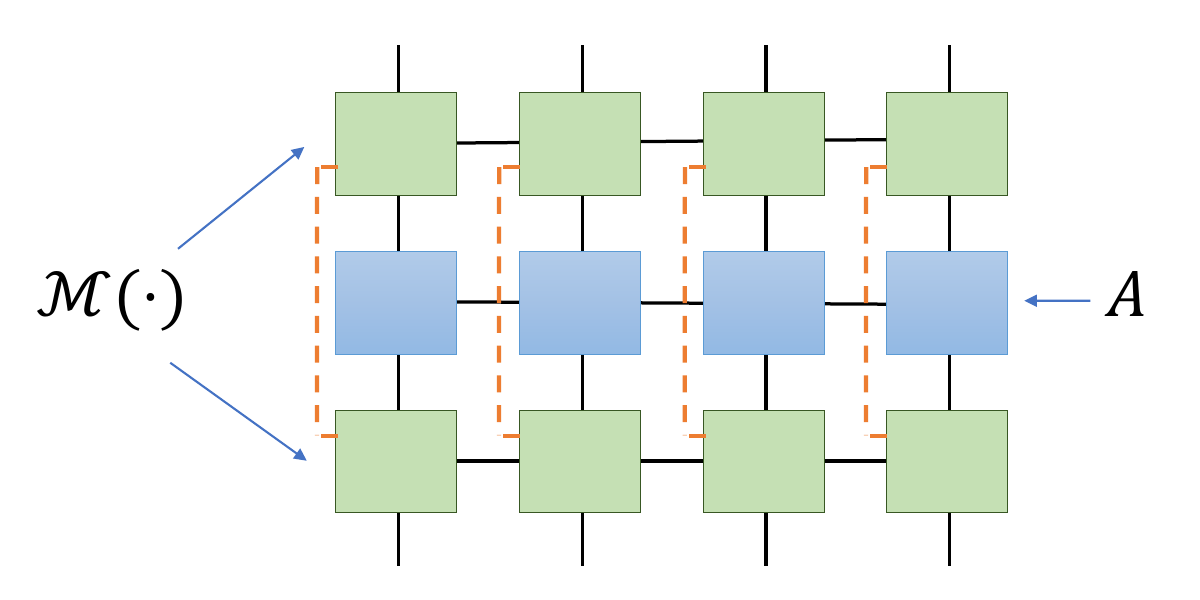}.
%\caption{Write some caption here}\label{visina8}
\end{figure}
That is, it can be written as a product of PEPOs acting on each of the two physical indices of $A$, and with an additional ``physical" index (in the figure, orange and dashed) that is contracted, such that $\mathcal{M}(A)=\sum_i M_D^{(i)} A (M_D^{(i)})^{\dagger}$ for some PEPOs $M_D^{(i)}$.

%%%%%%%%%%%%%%%%%%%%%
%%%%%%%%%%%%%%%%%%%%%%

\section{Local approximations to thermal states}\label{sec:thermal}

We start with local approximations to thermal states. Let $\mathcal{R}$ be a small region as defined in Sec. \ref{sec:setting}. The local thermal state is then $\text{Tr}_{\Lambda \setminus  \mathcal{R}}[\frac{e^{-\beta H}}{Z}]$, and the marginal of the PEPO approximation $\tilde{\rho}_k$ is $\text{Tr}_{\Lambda \setminus  \mathcal{R}}[\tilde{\rho}_k]$.

A first idea to approximate the marginals could be to simply consider a product $\bigotimes_i \text{Tr}_{\Lambda \setminus  \mathcal{R}_i}[\frac{e^{-\beta H}}{Z}]$ for some choice of adjacent regions $\{\mathcal{R}_i\}$. However, this clearly yields a large error in the regions that lie within two adjacent $\mathcal{R}_i$. Since we want to approximate all of them at once, we need a scheme that has no preferred partition of the lattice $\{\mathcal{R}_i\}$. This is possible with a uniform average over large enough partitions, and aided by local indistinguishability, which states that the marginal of large thermal states cna be approximated by the marginals of much smaller thermal states (see Lemma \ref{lem:locglob} for the precise statement).

One key assumption we need is the \emph{decay of correlations}
	\begin{equation}\label{eq:corrlength}
		\max_{X,Y} \frac{\left| \langle X\otimes Y \rangle_\beta-\langle X \rangle_\beta\langle Y \rangle_\beta \right|}{\vert\vert X \vert \vert \, \vert\vert Y \vert\vert \,} \le  \varepsilon(l),
	\end{equation}
where the optimization is over arbitrary observables separated by a distance $l$ on the lattice, and $\varepsilon(l)$ is a function decreasing with $l$. 

The most prominent decay is exponential $\varepsilon(l) \le e^{-l/\xi}$, which defines the thermal correlation length $\xi$.
This has been proven for translation invariant chains \cite{Araki69,bluhm2021exponential}, and there is strong evidence that it holds in any 1d thermal state \cite{Harrow_2020,bluhm2021exponential}. In higher dimensions it only holds above a finite threshold temperature $\beta^*$, as shown in \cite{Kliesch_2014,Fr_hlich_2015}, which also give a bound on the correlation length. However, we can also consider the polynomial decay $\varepsilon(l) \le \frac{R}{l^{d+1}}$ for some constant $R>0$. This might correspond to the behaviour at certain thermal phase transitions.

The decay of correlations is required to simulate local properties without a system size dependence, since one needs to be able to isolate them from distant regions. This is only possible if the correlations decay sufficiently fast, so that regions of width $ \sim \mathcal{O}(\xi)$ can be approximated independently of the rest. 

With this, we now show the main result of this section.
\begin{result}\label{th:thermal}
There is an explicit construction of a PEPO $\tilde{\rho}_k$ such that for any region $\mathcal{R}$ on the lattice $\Lambda$ it locally approximates the thermal state 
\begin{equation}
    \vert \vert \text{\normalfont{Tr}}_{\Lambda\setminus \mathcal{R}}[\frac{e^{-\beta H}}{Z}]-\text{\normalfont{Tr}}_{\Lambda\setminus \mathcal{R}}[\tilde{\rho}_k]\vert \vert_1 \le \epsilon.
\end{equation}
The bond dimension is bounded as follows.
\begin{itemize}[leftmargin=*]
    \item For $d=1$, assuming correlations decay exponentially, it is a MPO with bond dimension
\begin{align}\label{eq:thermalD1}
    D \le &\left(\frac{k+\xi}{\epsilon}\right)\\ & \times \exp\left[{\tilde{\mathcal{O}}\left( \max\{\beta, \sqrt{\beta \log(\frac{k+\xi}{\epsilon^2})}\}\right)}\right],\nonumber 
\end{align}
which is quasilinear in $(k+\xi)/\epsilon$ for any $\beta \simeq \mathcal{O}(1)$.
\item In higher dimensions $d>1$ and at high temperature $\beta \le \beta^* \equiv \log\left((1+\sqrt{1+4/\gamma})/2 \right)/2h  $ the bond dimension is
\begin{equation}\label{eq:thermalD2}
D\le \left(C' \beta d \max \left\{\frac{k^d} {\epsilon^{2}},\frac{d^{2d} \xi(\beta)^{d^2}}{\epsilon^{d+1} }\right\} \right)^{\mathcal{O}\left(\beta d \right)},
\end{equation}
where $C'$ is constant and the correlation length is $\xi(\beta) \equiv \left\vert \left(\log [\gamma e^{2\beta h}(e^{2\beta h}-1)] \right)^{-1}\right\vert$.
\item For lower temperatures, if the correlations decay polynomially $\varepsilon(l) \le \frac{R}{l^{d+1}}$, $D$ is bounded as in Eq.\eqref{eq:thermalD1}, \eqref{eq:thermalD2} but replacing $\xi^d$ with $R$.
\end{itemize}

\end{result}

\begin{figure}[t]
    \includegraphics[width=0.7\linewidth]{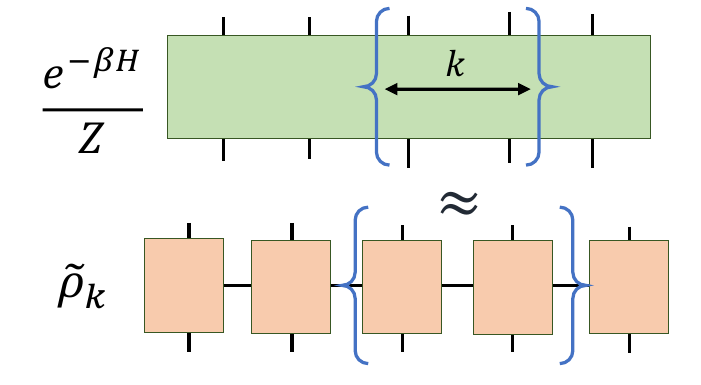}
    \caption{Schematic illustration of Result \ref{th:thermal}, where we approximate subsystems of a thermal state. $\tilde \rho_k$ is the local PEPO approximation, and the regions approximated are of length $k$.}
    \label{fig:result1}
\end{figure}

The bond dimension in both cases grows with $k$, $\xi$, $\beta$ and $\epsilon^{-1}$, as expected.
This result implies a good approximation in any local expectation value, since $\vert \vert \text{Tr}_{\Lambda\setminus \mathcal{R}}[\rho]-\text{Tr}_{\Lambda\setminus \mathcal{R}}[\sigma]\vert \vert_1= \max_{\text{supp}(A) \in \mathcal{R} ,\norm{A}=1} \tr{A(\rho-\sigma)}$. See Fig. \ref{fig:result1} for an illustration.

The proof is shown in Appendix \ref{app:localthermal}. It draws inspiration from previous results on local approximations of pure states \cite{huang2015computing,schuch2017matrix,Dalzell_2019,huang2019approximating}. The PEPO here is the uniform average of tensor products of approximations to local thermal states. These lie on consecutive hypercubes of a given size $l_0^d$, which span the whole lattice. The average is taken over all the different $l_0^d$ partitions of the lattice (that is, the different displacements of a given partition into hypercubes, see Fig. \ref{fig:2DThermal}). 

Each of the PEPOs on the hypercubes approximates $\text{Tr}_{\Lambda \setminus  \mathcal{R}}[\frac{e^{-\beta H}}{Z}]$ well for a given partition provided that $\mathcal{R}$ is far from the boundary between adjacent hypercubes. That this happens for most partitions is guaranteed by a result from \cite{brandao2019finite}, which shows that any local marginal of a thermal state does not depend on the regions far away from it (here how ``far away'' is determined by the thermal correlation length). This is exactly the idea behind local indistinguishability \cite{brandao2019finite}, and the related concept of \emph{locality of temperature} \cite{Hartmann_2006,Ferraro_2012,Kliesch_2014,Hern_ndez_Santana_2015,santana2020locality}, which states that $i)$ subsystems of thermal states of quantum Hamiltonians are robust to distant perturbations and that $ii)$ they are close to the marginals of the thermal state of their vicinity (see \cite{Kliesch_2014} for further discussions on this idea). 

The smaller PEPOs within the hypercubes can then be taken to be any of the existing global approximations to thermal states. The current best estimates are given in \cite{kuwahara2020improved} for 1d, which is $D\le \exp(\mathcal{O}\left(\sqrt{\beta \log (l_0/\epsilon)}\right) )$ and in \cite{Molnar_2015} for higher dimensions, which is $D\le \left(\frac{\beta l_0^d}{\epsilon}\right)^{\mathcal{O}(\beta d)}$. In the proof, one has to choose $l_0$ small to keep the bond dimension controlled, but still large enough such that the error from the local indistinguishability estimate is $\mathcal{O}(\epsilon)$. This leads to Eq. \eqref{eq:thermalD1} (by choosing $l_0 \propto \frac{k+\xi}{\epsilon}$) and Eq. \eqref{eq:thermalD2} (by choosing $l_0 \propto \max \{k \epsilon^{-d},\frac{d \xi^d}{\epsilon} \}$).

Let us comment on the algorithmic implications of Result \ref{th:thermal}. To build the MPO/PEPO here one just needs to construct them as given by the prescriptions of \cite{kuwahara2020improved} and \cite{Molnar_2015} respectively. In \cite{kuwahara2020improved}, the 1-dimensional approximation is defined as a product of the Taylor expansion of operators $e^{-\beta H_j}$, where $H_j$ is the Hamiltonian in a small region. Similar algorithms have already appeared in the literature \cite{Chen_2017,Chen_2018}. In \cite{Molnar_2015}, the higher dimensional approximation is based on the linked cluster expansion \cite{Hastings_2006}, which can in principle also be implemented numerically \cite{Rigol_2006,Vanhecke_2021}. Standard MPO/PEPO results \cite{Schollw_ck_2011,Pirvu_2010} guarantee that these approximations can be computed via an algorithm with run-time $\text{poly}\left( D, N \right)$.

%%%%%%%%%%%%%%%
%%%%%%%%%%%%%%%

\section{Local approximations to time evolution}\label{sec:timeevol}

We now focus on efficient approximations to the Heisenberg time evolution $e^{-itH}Ae^{itH}$. The existing results on global approximations \cite{Osborne_2006,kuwahara2020improved,Molnar_2015,Kliesch_2014} show that the bond dimension of a PEPO $M^t$ that approximates $\norm{e^{-itH}-M^t}\le \epsilon$ must grow with system size. We again drop this dependence when our target is the Heisenberg evolution of local operators. The key idea is to use the Lieb-Robinson bound \cite{lieb1972finite}, which states that the evolution $e^{-itH}Ae^{itH}$ is restricted to a certain ``light-cone" much smaller than the whole system. The evolution in this light-cone can be though of as generated by the Hamiltonian $H$ restricted to the vicinity of $A$.

To simulate this evolution, the first idea could be to simply reduce the problem to simulating the Lieb-Robinson light-cone exactly, which only requires a unitary in a region of size $\propto \mathcal{O}(v_{\text{LR}}t+\log(\epsilon^{-1}))$, with $v_{\text{LR}}$ is the Lieb-Robinson velocity. While this can be done with a bond dimension independent of system size, as previously pointed in \cite{Hastings_2008}, it would only be a good approximation for observables in a specific small region.

Here we show how the idea of using the effective light-cones of the evolution can be pushed further in order to build a single tensor network that approximates the Lieb-Robinson light-cone of any local operator. The statements for one and higher dimensions differ significantly, and are presented separately. 

%%%%%%%%%%%%%%
%%%%%%%%%%%%%%

\subsection{One dimension}
In one dimension, our main result essentially shows that the previous global approximation scheme from \cite{Osborne_2006} simulates well any Lieb-Robinson lightcone, and that in order to do so one only requires a bond dimension depending on the size of the lightcone, and not on $N$. The result is as follows.

\begin{figure}[t]
    \includegraphics[width=0.95\linewidth]{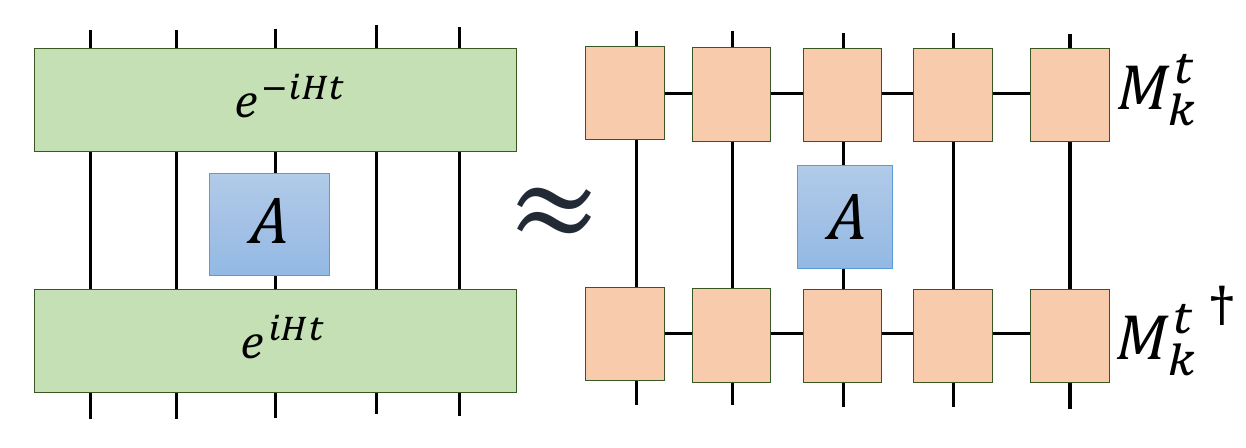}
    \caption{Schematic illustration of Result \ref{th:1Dtime}, where we approximate the Heisenberg evolution $e^{-iHt} A e^{iHt}$ within the effective light-cone. The operator $A$ can have support on at most $k$ adjacent sites.}
    \label{fig:result2}
\end{figure}

\begin{result}\label{th:1Dtime}
For any operator $A$ with support on a small region $\mathcal{R}$, the Heisenberg time evolution is well approximated as
\begin{equation}
\vert \vert e^{-itH}Ae^{itH} -  M^{t}_{k}A (M^{t})^ \dagger_{k} \vert \vert \le 3\epsilon \vert \vert A \vert \vert, 
\end{equation}
where $ M^{t}_{k}$ is an MPO with bond dimension
\begin{equation}\label{eq:1DTBD}
    D \le e^{\mathcal{O}\left( \vert t \vert\right)}\times \text{\normalfont{poly}} \left(\frac{k+v_{\text{LR}}t+\log\frac{1}{\epsilon}}{\epsilon} \right).
\end{equation}
\end{result}
That is, the bond dimension scales polynomially in $k,\epsilon^{-1}$ with a constant degree, and exponentially in time, which is consistent with the expected linear growth in entanglement along a generic time evolution \cite{Bravyi_2007,Eisert_2006,marien2016entanglement}. The numerator in the polynomial of Eq. \eqref{eq:1DTBD} corresponds to the size of the light-cone that needs to be approximated.  See Fig. \ref{fig:result2} for an illustration.

The proof is shown in Appendix \ref{app:1DTime}.
The construction is the same as that of \cite{Osborne_2006}, which shows that $e^{-itH}$ can be approximated by a quantum circuit of depth two, in which the size of the gates grows with $t, \epsilon^{-1}$ and system size $N$. To drop the system size dependence we give an argument based on the Lieb-Robinson bound which shows that to simulate the ligh-cone of $A$ one just needs to approximate the effective region of size $\propto \mathcal{O}(v_{\text{LR}}t+\log(\epsilon^{-1}))$. The important point is that this can be done such that the same MPO simulates the light-cone of \emph{any} local operator. For the argument to hold, it is crucial that the MPO of \cite{Osborne_2006} is a depth-$2$ quantum circuit, which is impossible in higher dimensions \cite{Haah_2021}. 

This MPO can also be implemented in practice \cite{Hastings_2008}, following the explicit construction of \cite{Osborne_2006}, which consists on the subsequent application of two local Hamiltonian evolutions. Thus, Result \ref{th:1Dtime} also guarantees an efficient 1d algorithm for short times. The result naturally extends to the simulation of any extensive sum of local operators by linearity. 

The exponential scaling in time originates from the error in the Lieb-Robinson bound. For systems with different lightcones, the growth in the bond dimension may be much smaller. For instance, many-body localized systems with a ``zero velocity" Lieb-Robinson bound $\propto t e^{-l}$ \cite{Burrell_2007,Hamza_2012} can instead be approximated with a bond dimension growing polynomially in time \cite{Hastings_2008}.

%%%%%%%%%%%%%%%%%%
%%%%%%%%%%%%%%%%%%

\subsection{Higher dimensions}

For higher dimensions, we resort to the idea in Sec. \ref{sec:thermal} of using partitions of the lattice into hypercubes of length $l_0$. The simple approach taken here is to first show that $e^{-itH}Ae^{itH}$ is close to the evolution of an effective Hamiltonian in the hypercube, and then approximate that effective evolution with a PEPO of small bond dimension. This PEPO can be constructed with the cluster expansion results from \cite{Molnar_2015,Kliesch_2014}, adapted to real time evolution (as described in Appendix \ref{app:clustertime}). The bond dimension of each of these is $D_{l_0} \le \exp \left({\mathcal{O}\left(\vert t \vert \log{\frac{\vert t\vert  l_0^d}{\epsilon}}\right)}\right)$.

The approximation is then accurate if $i)$ the hypercube is large enough and $ii)$ $A$ is sufficiently far away from the boundary between hypercubes. However, to construct a scheme that applies to local operators $A$ in any region, we need a tensor network that implements the approximation of a hypercubes \emph{conditioned} on the support of $A$. This conditioning requires something more involved than just acting with a single PEPO and its adjoint: a tensor map as described in Sec. \ref{sec:setting}. In Appendix \ref{app:highdtime} we construct a map that implements a PEPO $M_p$ among a possible list as $M_p A M_p^{\dagger}$, where $M_p$ is determined by the region $
\mathcal{R}$ that supports $A$. Since the map is linear, it can also act on extensive sums of local operators.

Given that there are $l_0^d$ possible partitions into hypercubes, we show that this can be done with a tensor network of bond dimension $\mathcal{O}(l_0^{3d} D^2_{l_0})$.   Choosing $l_0 =\tilde{\mathcal{O}}
\left(k+v_{\text{LR}}t+\log{\frac{1}{\epsilon}} \right)$ then yields the following result. 

\begin{result}\label{th:timehighd}
There is an explicit construction of a linear PEPO map $ \mathcal{M}^{t}_{k}( \cdot)$ such that, for any operator $A$ with support on a small region $\mathcal{R}$, the Heisenberg time evolution is well approximated as
\begin{equation}
\vert \vert e^{-itH}Ae^{itH} -  \mathcal{M}^{t}_{k}(A) \vert \vert \le 3\epsilon \vert \vert A \vert \vert, 
\end{equation}
If $A$ is a product of Pauli matrices, the map has bond dimension 
\begin{equation}
    D \le  \mathcal{O}\left(d \vert t \vert\frac{k+v_{\text{LR}}t+\log{1/\epsilon}}{\epsilon}  \right)^{\mathcal{O}(\vert t \vert d )},
\end{equation}
whereas for arbitrary $A$ the bond dimension is
\begin{equation}
    D \le  \mathcal{O}\left(d \vert t \vert 4^k\frac{v_{\text{LR}}t+k(1+\log{4/\epsilon)}}{\epsilon}  \right)^{\mathcal{O}(\vert t \vert d )}.
\end{equation}
\end{result}

The proof is shown in \ref{app:highdtime}, where we also explain how to construct the tensor network that applies a given PEPO conditioned on the support of the input $A$. This construction of a tensor network map that acts differently depending on some feature of the input (here, the non-trivial support of the observable) has, as far as we know, not previously appeared in the literature. We believe that this or similar schemes have the potential for further applications.

%%%%%%%%%%%%%%%%%
%%%%%%%%%%%%%%%%

\section{Applications}\label{sec:appl}

Our results apply to a wide range of physical situations. For instance, Result \ref{th:thermal} directly shows that it is possible to compute local thermal averages of arbitrary local operators, such as the energy or average magnetization. They can, however, also be used to guarantee approximations of more complex objects. We now elaborate on two of them. 

%%%%%%%%%%
%%%%%%%%%%

\subsection{Correlation functions}

A quantity that appears in many relevant situations, mostly pertaining to linear response theory \cite{Kubo57}, is the 2-point correlation function 
\begin{equation} \label{eq:autocorrelation}
    \langle A(t) A \rangle_\beta = \tr{\frac{e^{-\beta H}}{Z} e^{-itH}Ae^{itH} A },
\end{equation}
where $A$ is often taken to be an extensive operator. By this, we mean that it has uniform support throughout the lattice as $A=\frac{1}{N}\sum_x A_x \otimes \mathbb{I}_{\lambda \setminus x}$, where each term has bounded norm $\norm{A_x}\le \norm{A} $ and acts on at most $k$ consecutive sites. This is the central object in various areas of quantum dynamics, such as the study of quantum transport, for which $A$ is taken to be a current operator \cite{bertini2020finitetemperature} of the relevant conserved quantities. 

In the next result, we show how our constructions serve to approximate it. We focus in the one dimensional case, for which we again assume that Eq. \eqref{eq:corrlength} holds with an exponential decay.  In higher dimensions Eq. \eqref{eq:autocorrelation} involves a contraction which is typically computationally hard \cite{Schuch_2007,Haferkamp_2020} (although this is potentially not a problem in physically relevant contexts, see \cite{Schwarz_2017}). The result is as follows.

\begin{result}\label{th:autocorr}
The correlation function of an extensive observable $A$ in one dimension can be approximated as
\begin{equation}\label{eq:finalautocorr}
  \vert \tr{\tilde{\rho}_{k'} M_{k}^t A (M_k^{t})^{ \dagger} A} -   \langle A(t) A \rangle_\beta \vert \le 12 \epsilon \norm{A}^2,
\end{equation}
where $k' = \mathcal{O}\left( \xi \log{\frac{1}{\epsilon}}+v_{\text{LR}}t+k\right)$, and the operators  $\tilde{\rho}_{k'}$ and $M_{k}^t$ are as defined in Results \ref{th:thermal} and \ref{th:1Dtime}.
\end{result}
The proof is shown in Appendix \ref{app:autocorr}. In simple terms, Result \ref{th:autocorr} implies that $\langle A(t) A \rangle_\beta$ can well be approximated by a contraction of MPOs of bond dimension at most
\begin{equation}
    D \le e^{\mathcal{O}\left( \vert t \vert\right)}\times \text{\normalfont{poly}} \left(\frac{k+v_{\text{LR}}t+\xi \log\frac{1}{\epsilon}}{\epsilon} \right).
\end{equation}

  \begin{figure}[t]
    \includegraphics[width=0.95\linewidth]{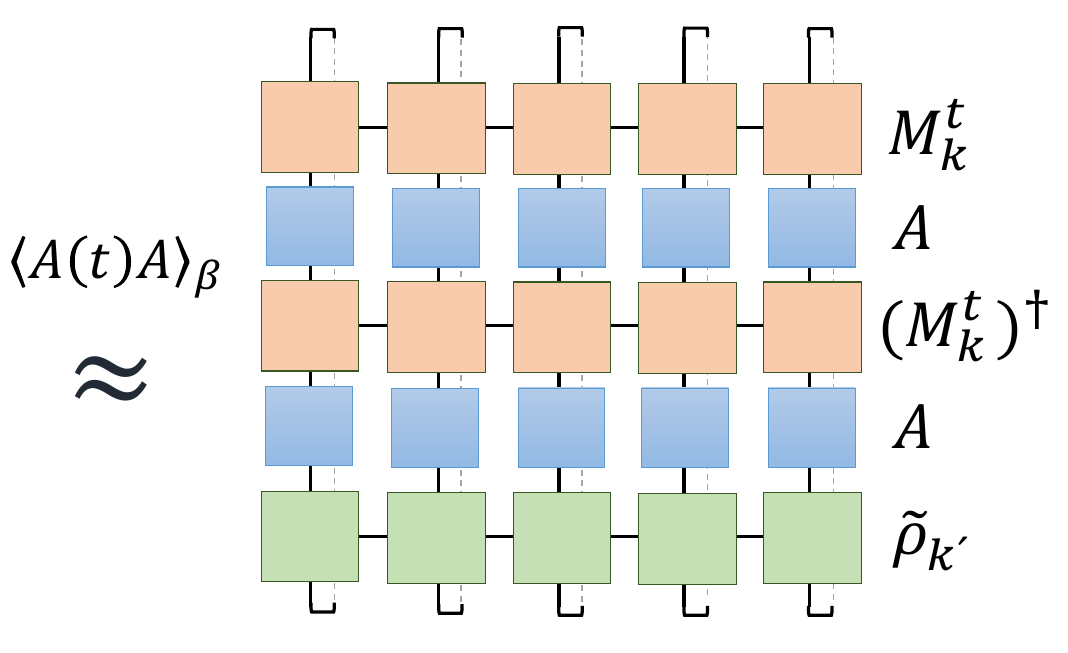}
    \caption{Schematic illustration of Result \ref{th:autocorr} with the contraction of tensors involved in approximating the correlation functions $\langle A(t) A \rangle_\beta$, where $A$ may be an extensive observable.}
    \label{fig:corrfunc}
\end{figure}
See Fig. \ref{fig:corrfunc} for an illustration.
The biggest drawback of this result is the fast growth in time $t$, which is nevertheless expected in general. This does not necessarily prevent the algorithms from reaching interesting timescales \cite{Alhambra_2020}, and is likely an overestimation for many important situations at late times. Nevertheless, we believe that this result mathematically justifies the success of previous tensor network approaches to computing correlation functions \cite{Barthel_2009,Karrasch_2012,Karrasch_2013,Barthel_2013,Tiegel_2014,Karrasch_2015,bertini2020finitetemperature}.

%%%%%%%%%%%%%%%
%%%%%%%%%%%%%%%

\subsection{Quantum quenches}

In quantum quenches, one starts with a pure initial state $\ket{\Phi}$. This is often an easy-to-prepare state, such as a ground state of a gapped model, or a product state. Then, the Hamiltonian is suddenly switched to some arbitrary $H$, and the subsequent time evolution is tracked through expectation values of local observables $\langle A(t) \rangle\equiv \bra{\Psi} A(t) \ket{\Psi}$. This time evolution can be simulated with the results in Sec. \ref{sec:timeevol}, which give an upper bound on the bond dimension required (e.g. Eq. \eqref{eq:1DTBD}). 

This upper bound, however, grows very fast with time, and may become too large at relevant timescales such as the thermalization or Thouless times \cite{D_Alessio_2016}. This is so even if the local marginals of the evolved state are simply described by a thermal state or a GGE, which depends on very few parameters. In those cases, it is expected that local evolution at late times can also be approximated with a small bond dimension \cite{leviatan2017quantum,White_2018,Berta_2018,Xu_2019,Hallam_2019}. 

One of the main results of \cite{Dalzell_2019} (applicable in 1d) can help in this setting: the local marginals of \emph{any} state $\ket{\Phi(t)}$ on $k$ sites can be approximated with an MPS $\ket{\Psi}$ with bond dimension $D \le \exp\left(k/\epsilon \right)$. One can then potentially simulate $\langle A(t) \rangle$ in 1d with a bond dimension
\begin{equation}
    D \le \min \left \{ e^{\mathcal{O}\left( \vert t \vert\right)}\times \text{\normalfont{poly}} \left(\frac{k+v_{\text{LR}}t+\log\frac{1}{\epsilon}}{\epsilon} \right) ,e^{\frac{k}{\epsilon}}\right \}.
\end{equation}
It is not clear whether an efficient algorithm to find $\ket{\Psi}$ exists \cite{Dalzell_2019}. However, numerical schemes for approximating $\langle A(t) \rangle$ at long times with low bond dimension have already been devised \cite{leviatan2017quantum,White_2018}.

%%%%%%%%%%%%%%%%%%%%%
%%%%%%%%%%%%%%%%%%%%%

\section{Conclusions}

We have shown how to construct tensor network representations of thermal states and time evolution, in a way that local observables are well approximated. This allows us to provably achieve a bond dimension independent of system size in all cases, which contrasts with what is achieved by previous global approximations. These results have implications on the tensor network simulation of various equilibrium and out of equilibrium situations, and help to mathematically justify the success of previous numerical results.

Since they hold for any local Hamiltonian, and any timescale and temperature, they are likely not tight in particular cases of interest. For instance, when simulating the long time dynamics of a system that has already thermalized, it seems likely that only a much smaller bond dimension (perhaps independent of time) is required. This intuition is present in previous specific algorithms \cite{White_2018,leviatan2017quantum}.

In all of the proofs, except for Result \ref{th:1Dtime}, the constructions involve an average over ``local approximations", which we expect to be unnecessary in practice. This also applies to previous results for ground state approximations \cite{huang2015area,schuch2017matrix,Dalzell_2019,huang2019approximating}.  It would be interesting to find arguments to circumvent this proof idea, perhaps akin to the proof of Result \ref{th:1Dtime} or to other features such as the Markov property of thermal states \cite{Kato_2019,Kuwahara_2020}. For 1d, this question was dealt with in \cite{huang2021locally}, where it was shown that there exists an MPO that reproduces the local expectation values of the thermal state with bond dimension $D \le \exp{\tilde{\mathcal{O}}(\beta^{2/3}+\sqrt{\beta \log(k/\epsilon)})}$. This significantly improves on the result of Eq. \eqref{eq:thermalD1}, at the price of not giving an explicit construction of the MPO. It may also be possible to obtain a better higher dimensional generalization of Result \ref{th:1Dtime} that does not require a tensor map, as Result \ref{th:timehighd} does.

\acknowledgements

The authors acknowledge funding from the Alexander von Humboldt Foundation and from ERC Advanced Grant QUENOCOBA under the EU Horizon 2020 program (Grant Agreement No. 742102) and within the D-ACH Lead-Agency Agreement through project No. 414325145 (BEYOND C). 

\bibliography{references}

\widetext
\newpage
\appendix

%%%%%%%%%%%%%%%%
%%%%%%%%%%%%%%%

\section{Review of the cluster expansion approximation to thermal and real time evolution in arbitrary dimension}\label{app:clusterexp}

We briefly review the main result of \cite{Molnar_2015}, which provides PEPO approximations for global thermal states in arbitrary spatial dimensions via the cluster expansion \cite{Hastings_2006,Kliesch_2014}. We then explain how an analogous statement holds for the operator $e^{itH}$.

%%%%%%%%%%%%
%%%%%%%%%%%%

\subsection{Thermal state at any temperature} \label{app:thermalhighD}

Let us recall the definitions from Sec. \ref{sec:setting}: $H=\sum_x h_x$ is a $k$-local Hamiltonian on an arbitrary $d$-dimensional lattice, with up to $K \propto N$ terms, and such that $\max_x \vert \vert h_x \vert \vert \le h \propto \mathcal{O}(1)$, and the interaction graph has degree at most $z$. We also define the lattice growth constant as $\gamma$.

It was shown in \cite{Hastings_2006} that there exists an operator $\tilde{\rho}$, defined in terms of the cluster expansion (see \cite{Kliesch_2014} for a detailed proof), which is a good approximation to the thermal state for high temperatures. The statement is as follows. Let $\beta^*$ be a constant such that $\gamma e^{(2z-1)\beta^*h}(e^{\beta^* h}-1)<1$ (that is, $\beta^* \sim 1/h d$). If $\beta \le \beta^*$ , then
\begin{equation} \label{eq:approx1norm}
\vert \vert e^{-\beta H} -\tilde{\rho} \vert \vert_1 \le \vert \vert e^{-\beta H} \vert \vert_1 \left( \text{exp}\left( K \frac{x^L}{1-x} \right)-1 \right),
\end{equation}
where $x\equiv \gamma e^{(2z-1)\beta h}(e^{\beta h}-1)<1$. Here, $L$ is a free parameter that determines the size of the clusters in the approximation. Importantly, Eq. \ref{eq:approx1norm} also holds for the norm $\vert \vert ...\vert \vert_{2M}$ if we change the temperature to $\beta'=\frac{\beta}{2M}$, such that 
\begin{equation} \label{eq:approx2Mnorm}
\vert \vert e^{-\frac{\beta}{2M} H} -\tilde{\rho} \vert \vert_{2M} \le \vert \vert e^{-\frac{\beta}{2M} H} \vert \vert_{2M} \left( \text{exp}\left( K \frac{x'^L}{1-x'} \right)-1 \right),
\end{equation}
with $x'\equiv \gamma e^{(2z-1)\frac{\beta h}{2M}}(e^{\frac{\beta h}{2M}}-1)<1$. Then, Proposition $1$ in \cite{Molnar_2015} allows us to approximate the thermal state at any temperature.
\begin{proposition}\label{prop:prop1molnar}
If $\epsilon<1/3$ and
\begin{equation}
    \vert \vert e^{-\frac{\beta }{2M}H} - \tilde{\rho}  \vert \vert_{2M} \le \frac{\epsilon}{M}  \vert \vert e^{-\frac{\beta}{2M} H} \vert \vert_{2M}
\end{equation}
it follows that
\begin{equation}
     \vert \vert e^{-\beta H} - (\tilde{\rho} ^\dagger\tilde{\rho})^{2M}  \vert \vert_{1} \le \frac{\epsilon}{M}  \vert \vert e^{-\beta H} \vert \vert_{1}.
\end{equation}
\end{proposition}
To choose an $\epsilon$-close approximation in 1-norm at any temperature, we need to set $M$ large enough such that $\frac{\beta}{2M}\le \beta^* \sim\frac{1}{hd}$, which amounts to $M = \mathcal{O}(\beta h d)$. Then, for large $L$ the error in Eq. \eqref{eq:approx2Mnorm} is 
\begin{equation}
    \left( \text{exp}\left( K \frac{x'^L}{1-x'} \right)-1 \right) \simeq K x'^L (1-x')^{-1},
\end{equation}
where $x' =\mathcal{O}(1)$. That is, we need to set $L = \mathcal{O}\left(\log \frac{KM}{\epsilon}\right)=\mathcal{O}\left(\log \frac{\beta N d}{\epsilon}\right)$.

Moreover, it was shown in \cite{Molnar_2015} that $\tilde{\rho}$ is a PEPO with bond dimension $D \le e^{L}$, and thus $(\tilde{\rho} ^\dagger\tilde{\rho})^{2M}$ has $D \le e^{2M L}$.
We conclude that to achieve an error $\epsilon$ one requires a bond dimension
\begin{equation}
    D \le e ^{\mathcal{O}\left(\beta d \log{\frac{\beta d N}{\epsilon}}\right)}.
\end{equation}

%%%%%%%%%%%%%%
%%%%%%%%%%%%%%

\subsection{Real time evolution}\label{app:clustertime}
It can be easily seen that the proof of Eq. \eqref{eq:approx1norm} from \cite{Kliesch_2014}, and the cluster expansion analysis, also hold for an approximation of $e^{-itH}$ in operator norm, simply if one substitutes $\beta$ with $\vert t \vert$. This just follows from the observation that all the steps in the proof of \cite{Kliesch_2014} remain unchanged if one changes all the norms to operator norms, and that there is no further fundamental differences between the operators $e^{-\beta H}$ and $e^{-itH}$. In the same way, one also has the analogue of Proposition \ref{prop:prop1molnar} for operator norms, which allows us to extend the approximation for arbitrarily long times.

From this observations we conclude that there exists an PEPO $\tilde{U}_t$ with bond dimension $D$ such that
\begin{align}
    \vert \vert e^{-itH}-\tilde{U}_t \vert \vert \le \epsilon
\end{align}
and
\begin{equation} \label{eq:BDtimeev}
    D \le e ^{\mathcal{O}\left(\vert t \vert d \log{\frac{\vert t\vert d N}{\epsilon}}\right)}.
\end{equation}

%%%%%%%%
%%%%%%%%

\section{Local approximations to thermal states}\label{app:localthermal}

Here we show the main result regarding local approximations to thermal states. First, we need the following key assumption for the thermal state $\rho=e^{-\beta H}/Z$.

\begin{definition}[Clustering of correlations]
The state $\rho$ on a lattice system has $\epsilon(l)$-clustering of correlations if
	\begin{equation}
		\max_{X,Y} \frac{\left| \langle X\otimes Y \rangle_\beta-\langle X \rangle_\beta\langle Y \rangle_\beta \right|}{\vert\vert X \vert \vert \, \vert\vert Y \vert\vert \,} \le  \varepsilon(l),
	\end{equation}
	where $X$ has support of region $A$ only and $Y$ on region $B$ only, and $l\le \dist (A,B)$.
\end{definition}

The following result on local indistinguishability shows that marginals of thermal states when tracing out a big region are well approximated by the marginal of the thermal state of a much smaller lattice. This is the key ingredient to guarantee the faithfulness of local approximations.

\begin{lemma}\label{lem:locglob} [Theorem 4,  \cite{brandao2019finite}]
	Let H be a local bounded Hamiltonian, $\beta$ an inverse temperature and $\rho_{AB}=e^{-\beta H}/\tr{e^{-\beta H}}$. Let $AB_1B_2$ be a separation of the lattice such that $B_1$ shields $A$ from $B_2$ by a distance of at least $l$. Let $\rho_{AB_1}$ be the Gibbs state on region $AB_1$ only (that is, with the terms of the Hamiltonian $H$ that have support on $AB_1$ only). If the system has $\epsilon(l)$-clustering of correlations, then
	\begin{equation}
		\vert\vert \text{\normalfont{Tr}}_{B}[\rho_{AB}]-\text{\normalfont{Tr}}_{B_1}[\rho_{AB_1}] \vert \vert_1 \le C |\partial B_2|(\varepsilon(l/2)+c_1 e^{-c_2 l}),
	\end{equation}
	where $C>0$ and $c_1,c_2 > 0$ are constant and $|\partial B_2|$ is the size of the boundary between $B_1$ and $B_2$.
\end{lemma}
\noindent This lemma relies on the technique of \emph{quantum belief propagation} from \cite{Hastings_2007}, from which the contribution $ c_1 e^{- c_2 l}$ arises. See also Theorem 4 in \cite{Kliesch_2014} for a similar statement at high temperatures only.

\begin{figure}[t]
    \includegraphics[width=0.7\linewidth]{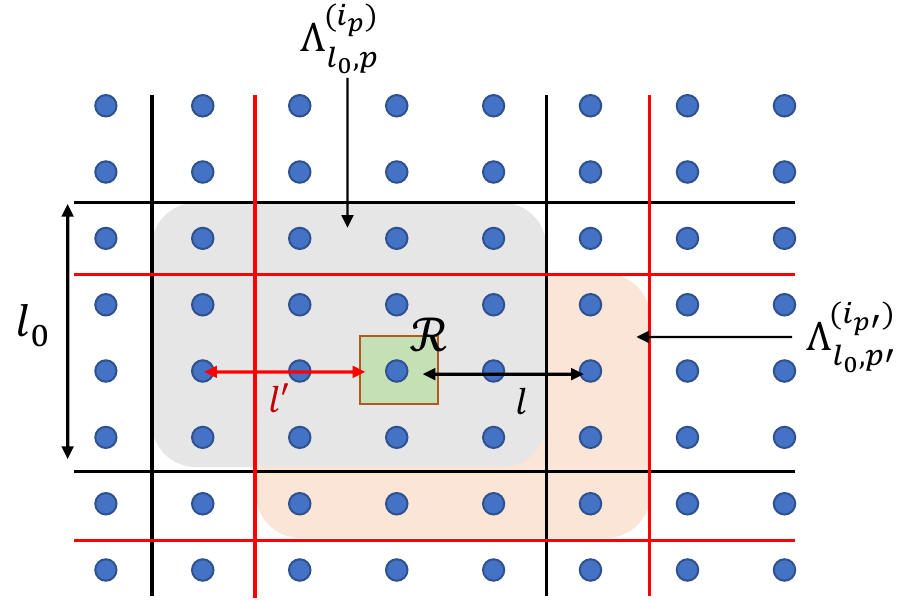}
    \caption{Schematic illustration of the idea behind the proof of the local approximations to thermal states. $\mathcal{R}$ is our region of interest, and the black and red grids represent two different partitions of the lattice, so that $\mathcal{R}$ is at a distance $l$ and $l'$ from each grid. The shaded regions correspond to the hypercubes (in this case, squares) $\Lambda^{(i)}_{l_0,p},\Lambda^{(i)}_{l_0,p'}$ of size $l_0^d$ on which $\mathcal{R}$ lies.}
    \label{fig:2DThermal}
\end{figure}

To prove the main result of the section, we need to first define the partitions $\mathcal{G}_p$ of the $d$-dimensional lattice $\Lambda$ into hypercubes of length $l_0$:
\begin{equation}
    \mathcal{G}_p = \{ \Lambda^{(i)}_{l_0,p}: \cup_i  \Lambda^{(i)}_{l_0,p} = \Lambda \}
\end{equation}
such that  $\vert \Lambda^{(i)}_{l_0,p} \vert \le l_0^d$. The parameter $p$ can for instance represent one of the corners of a given hypercube, which defines the position of the boundaries of all the hypercubes, and can adopt $l_0^d$ values. The hypercubes near the boundaries may have a smaller number of lattice sites. An illustration of these and the other elements of the proof is shown in Fig. \ref{fig:2DThermal}.

Let $\rho(\Lambda^{(i)}_{l_0,p})$ be the thermal state with Hamiltonian containing the terms of $H$ with support only on $\Lambda^{(i)}_{l_0,p}$.
Each $\rho(\Lambda^{(i)}_{l_0,p})$ can now be approximated to $\epsilon/3$ precision in $1$-norm by an PEPO $\tilde{\rho}(\Lambda^{(i)}_{l_0,p})$ of bond dimension $D_{l_0}$. In one dimension we have Theorem 3 in \cite{kuwahara2020improved}, which guarantees
\begin{equation}
   \log D_{l_0} \le \mathcal{O}\left( \max\{\beta, \sqrt{\beta \log(\frac{l_0}{\epsilon})}\}\times\log\left(\beta \log(\frac{l_0}{\epsilon})\right) \right),
\end{equation}
which is sub-polynomial in $l_0$. In higher dimensions, the result from \cite{Molnar_2015} described in Appendix \ref{app:thermalhighD} yields $D_{l_0}\le \exp{\mathcal{O}\left(\beta d \log \frac{\beta d l_0^d}{\epsilon}\right)}$. 

We also need that each of these PEPOs has trace $1$, which can be achieved with a small price in the precision as
\begin{align}\label{eq:traceone}
    \vert \vert \rho(\Lambda^{(i)}_{l_0,p})- \frac{\tilde{\rho}(\Lambda^{(i)}_{l_0,p})}{\tr{\tilde{\rho}(\Lambda^{(i)}_{l_0,p})}}\vert \vert_1 &\le
    \vert \vert \rho(\Lambda^{(i)}_{l_0,p})- \tilde{\rho}(\Lambda^{(i)}_{l_0,p})\vert \vert_1 + \vert \vert \tilde{\rho}(\Lambda^{(i)}_{l_0,p})- \frac{\tilde{\rho}(\Lambda^{(i)}_{l_0,p})}{\tr{\tilde{\rho}(\Lambda^{(i)}_{l_0,p})}}\vert \vert_1 
    \\
    & \le \frac{\epsilon}{3} +(1+\frac{\epsilon}{3}) \left \vert \left(1-\frac{1}{\tr{\tilde{\rho}(\Lambda^{(i)}_{l_0,p})}} \right) \right \vert \le \epsilon.
\end{align}
For simplicity, we will assume $\tr{\tilde{\rho}(\Lambda^{(i)}_{l_0,p})}=1$.

Now define the PEPO $\tilde{\rho}_{l_0,p}=\bigotimes_i \tilde{\rho}(\Lambda^{(i)}_{l_0,p})$. Our local PEPO approximation is the uniform average over all the partitions of the lattice into hypercubes
\begin{equation}\label{eq:defthermalMPO}
\tilde{\rho}=\frac{1}{l_0^d}\sum_p \tilde{\rho}_{l_0,p},
\end{equation}
which has bond dimension $D \le D_{l_0} \times l_0^{d}$. We now show this is a good approximation to the marginal of any region $\mathcal{R}\in \Lambda$ of maximum length $k < l_0$. First, we separate the terms in the sum over $p$ by whether $\mathcal{R}$ lies strictly inside one of the hypercubes $\Lambda^{(i)}_{l_0,p}$ or not,
\begin{align}
\vert \vert \text{Tr}_{\Lambda\setminus \mathcal{R}}[\rho]-\text{Tr}_{\Lambda\setminus \mathcal{R}}[\tilde{\rho}]\vert \vert_1 &\le \frac{1}{l_0^d}\sum_p \vert \vert \text{Tr}_{\Lambda\setminus \mathcal{R}}[\rho]-\text{Tr}_{\Lambda\setminus \mathcal{R}}[\tilde{\rho}_{l_0,p}]\vert \vert_1
\\ & \le \frac{1}{l_0^d}\left( \sum_{p: \mathcal{R} \nsubseteq \Lambda^{(i_p)}_{l_0,p} } \vert \vert \text{Tr}_{\Lambda\setminus \mathcal{R}}[\rho]-\text{Tr}_{\Lambda\setminus \mathcal{R}}[\tilde{\rho}_{l_0,p}]\vert \vert_1 +\sum_{p: \mathcal{R} \subset \Lambda^{(i_p)}_{l_0,p} } \vert \vert \text{Tr}_{\Lambda\setminus \mathcal{R}}[\rho]-\text{Tr}_{\Lambda\setminus \mathcal{R}}[\tilde{\rho}_{l_0,p}]\vert \vert_1 \right)
\\
& \le \frac{1}{l_0^d}\left( 2 k^d +\sum_{p: \mathcal{R} \subset \Lambda^{(i_p)}_{l_0,p} } \vert \vert \text{Tr}_{\Lambda\setminus \mathcal{R}}[\rho]-\text{Tr}_{\Lambda\setminus \mathcal{R}}[\tilde{\rho}_{l_0,p}]\vert \vert_1 \right),
\end{align}
where $i_p$ denotes the hypercube $\Lambda^{(i_p)}_{l_0,p}$ on which $\mathcal{R}$ lies, for a given $p$ (see Fig. \ref{fig:2DThermal}). Here, we just used the triangle inequality and the trivial bound $\vert \vert \rho-\sigma \vert \vert_1\le 2$.

The remaining terms can be bounded with Lemma \ref{lem:locglob}. For a given lattice partition $p$, let $l$ be the minimum distance between $\mathcal{R}$ and the boundary of $\Lambda^{(i_p)}_{l_0,p}$. Then,
\begin{align}
\vert \vert \text{Tr}_{\Lambda\setminus \mathcal{R}}[\rho]-\text{Tr}_{\Lambda\setminus \mathcal{R}}[\tilde{\rho}_{l_0,p}]\vert \vert_1 &\le \vert \vert \text{Tr}_{\Lambda\setminus \mathcal{R}}[\rho]-\text{Tr}_{\Lambda^{(i_p)}_{l_0,p}\setminus \mathcal{R}}[\rho(\Lambda^{(i)}_{l_0,p})]\vert \vert_1 +\vert \vert \text{Tr}_{\Lambda^{(i_p)}_{l_0,p}\setminus \mathcal{R}}[\rho(\Lambda^{(i)}_{l_0,p})]-\text{Tr}_{\Lambda\setminus \mathcal{R}}[\tilde{\rho}_{l_0,p}]\vert \vert_1 
\\ & =
\vert \vert \text{Tr}_{\Lambda\setminus \mathcal{R}}[\rho]-\text{Tr}_{\Lambda^{(i_p)}_{l_0,p}\setminus \mathcal{R}}[\rho(\Lambda^{(i)}_{l_0,p})]\vert \vert_1 +\vert \vert \text{Tr}_{\Lambda^{(i_p)}_{l_0,p}\setminus \mathcal{R}}[\rho(\Lambda^{(i)}_{l_0,p})]-\text{Tr}_{\Lambda^{(i_p)}_{l_0,p}\setminus \mathcal{R}}[\tilde{\rho}(\Lambda^{(i)}_{l_0,p})]\vert \vert_1 
\\ & \le 2Cd l_0^{d-1}(\varepsilon(l/2)+c_1 e^{-c_2 l}) + \frac{\epsilon}{3}
\end{align}
where in the first line we used the triangle inequality, in the second the fact that the PEPOs have trace $1$ by assumption and in the third Lemma \ref{lem:locglob} (with $\vert \partial B_2 \vert=2dl_0^{d-1}$) and the definition of $\epsilon$. Given that the area of a hypercube of edge $l$ is $2d l^{d-1}$,

\begin{align}
\vert \vert \text{Tr}_{\Lambda\setminus \mathcal{R}}[\rho]-\text{Tr}_{\Lambda\setminus \mathcal{R}}[\tilde{\rho}]\vert \vert_1 &\le
 \frac{2 k^d}{l_0^d} + \frac{\epsilon}{3} +  2Cd l_0^{-1}\sum_{p: l=\text{dist}(\mathcal{R},\partial \Lambda^{(i)}_{l_0,p}) }(\varepsilon(l/2)+c_1 e^{-c_2 l})
 \\& \le \frac{2 k^d}{l_0^d} + \frac{\epsilon}{3} + 4d^2 C l_0^{-1}\sum_{l=0}^{(l_0-k)/2}  l^{d-1} (\varepsilon(l/2)+c_1 e^{-c_2 l}).\label{eq:proofsum}
\end{align}

We now focus on the two different types of decay of the function $\varepsilon(l/2)$: exponential and polynomial.

\noindent \textbf{Exponential:} In this case $\varepsilon(l/2) \le e^{-l /2 \xi}$.  This has been shown in 1d translation invariant chains \cite{Araki69,bluhm2021exponential}, and it is believed to hold for all 1d systems \cite{Harrow_2020,bluhm2021exponential}. It has also been shown for higher dimensional systems at high temperature using the cluster expansion technique in \cite{Kliesch_2014}.
There, it is shown that, given $\beta^*\equiv \log\left((1+\sqrt{1+4/\gamma})/2 \right)/2h$, for every $\beta < \beta^*$, it holds that
\begin{equation}
\varepsilon(l) \le \frac{4 \min\{\vert \partial A \vert,\vert \partial B \vert\}}{\log3 (1-e^{-1/\xi(\beta)})}e^{-l/\xi(\beta)},
\end{equation}
where the correlation length is $\xi(\beta)=\left\vert \left(\log [\gamma e^{2\beta h}(e^{2\beta h}-1)] \right)^{-1}\right\vert$.

Starting from the sum in \eqref{eq:proofsum}, then
\begin{equation}
    \sum_{l=0}^{(l_0-k)/2}  l^{d-1} (\varepsilon(l/2)+c_1 e^{-c_2 l})\le (c_1+1)\sum_{l=0}^{\infty}  l^{d-1} \text{exp}\left(-\frac{l}{(c_2+2) \xi}\right) =\mathcal{O}(\xi^d).
\end{equation}
Choosing $l_0 = C' \max \{k \epsilon^{-d},\frac{d^2 \xi^d}{\epsilon} \}$ for some constant $C'>0$, we obtain $\vert \vert \text{Tr}_{\Lambda\setminus \mathcal{R}}[\rho]-\text{Tr}_{\Lambda\setminus \mathcal{R}}[\tilde{\rho}]\vert \vert_1 \le \epsilon$.
The bond dimension in 1d is hence
\begin{equation}\label{eq:ap1D}
    D \le \left(\frac{k+\xi}{\epsilon}\right)\times \exp\left[{\mathcal{O}\left( \max\{\beta, \sqrt{\beta \log(\frac{k+\xi}{\epsilon^2})}\}\times \log\left(\beta \log(\frac{k+\xi}{\epsilon^2}) \right)\right)}\right], 
\end{equation}
which for $\beta = \mathcal{O}(1)$ is quasilinear in $(k+\xi)/\epsilon$, $ D \le C' \left(\frac{(k+\xi)}{\epsilon}\right)^{1+o(1)}$. In higher dimensions, it is 
\begin{equation}\label{eq:apHD}
D\le \left(C' \beta d \max \{\frac{k^d} {\epsilon^{2}},\frac{d^{2d} \xi^{d^2}}{\epsilon^{d+1} }\} \right)^{\mathcal{O}\left(\beta d \right)},
\end{equation}
which is polynomial in $k$ and $\epsilon^{-1}$.

\noindent \textbf{Polynomial:} Now assume $\varepsilon(l/2) \le \frac{R}{l^{d+1}}$ for any $d$ (note that this assumption is likely not necessary in 1d). Then the following sum converges
\begin{equation}
    \sum_{l=0}^{(l_0-k)/2}  l^{d-1} (\varepsilon(l/2)+c_1 e^{-c_2 l})\le (R +c_1)\sum_{l=0}^{\infty}  l^{d-1} \frac{1}{l^{d+1}} =\mathcal{O}(R),
\end{equation}
so we simply need to choose $l_0 = C' \max \{k \epsilon^{-d},\frac{d R}{\epsilon} \}$ for some constant $C'>0$. The resulting bond dimension is thus the same as Eq. \eqref{eq:ap1D}, \eqref{eq:apHD} with $\epsilon^d$ replaced with R.

%%%%%%
%%%%%%

\section{Local approximations to time evolution}
Unlike for thermal states, the proofs and statements for one and higher dimensions differ significantly, and are presented separately.
\subsection{One dimension}\label{app:1DTime}
This is based on the main result of \cite{Osborne_2006}, and also relies on the Lieb-Robinson bound \cite{lieb1972finite,Bravyi_2006}. 

\begin{lemma}[Lieb-Robinson bound]\label{le:Liebrobinson}
Let $A$ be a local observable on $k$ sites and $H=
\sum_x h_x$ a uniformly bounded Hamiltonian with finite interaction range. Then there exist constants $c,v_{\text{LR}}\geq 0$ such that for all $X$ with $l:=\text{dist}(A,X^c)\geq 2d-1$ we have
\begin{align}
    \norm{e^{-itH}Ae^{itH} - e^{-itH_X}Ae^{itH_X}}\leq \norm{A}c  l^{d-1}\e^{v_{\text{LR}} t - l},
\end{align}
where $H_X$ contains only the terms of $H$ away from $A$ by a distance at most $l$. That is
\begin{align}
    H_X &= \sum_{x: \mathrm{supp}(h_x)\subseteq X} h_x.
\end{align}
\end{lemma}

\begin{figure}[t]
    \includegraphics[width=0.8\linewidth]{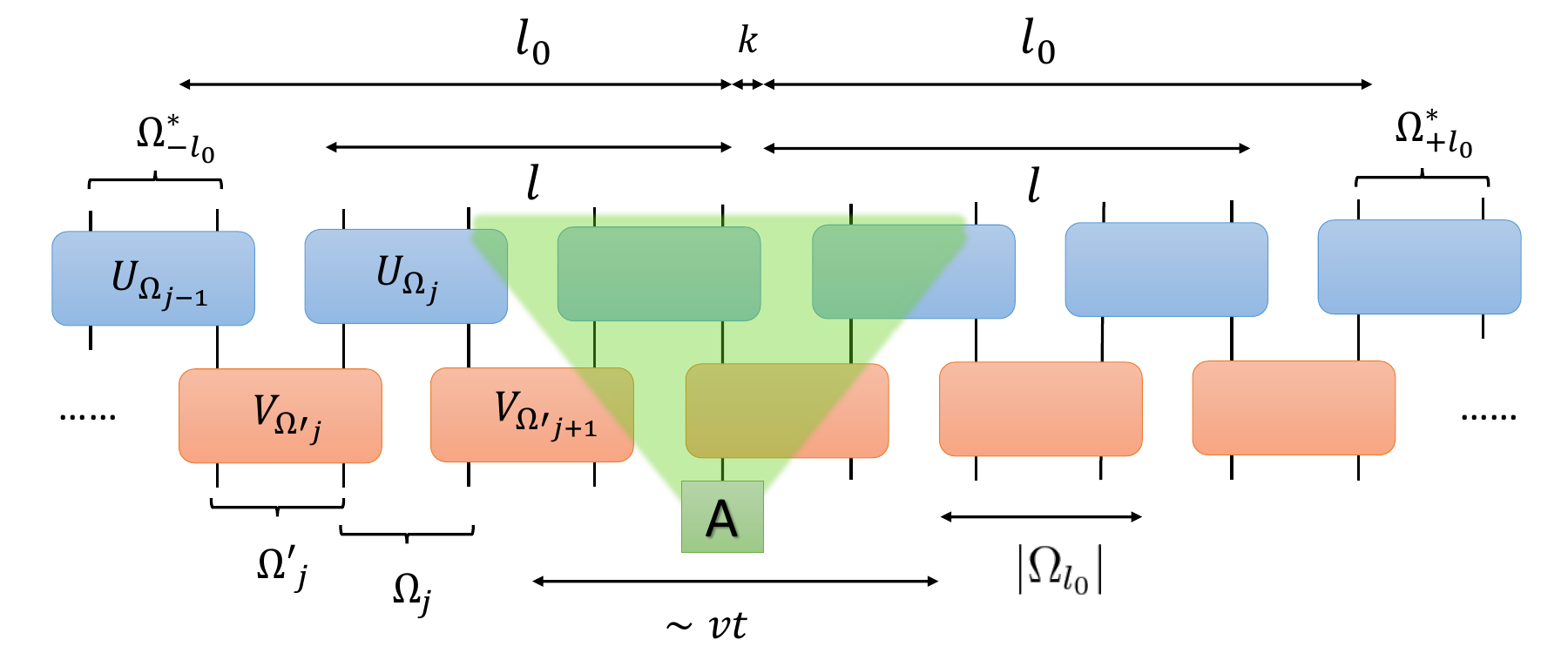}
    \caption{Schematic illustration of the proof of the local approximation to time evolution. The green area represents the effective Lieb-Robinson lightcone of $A$, and the $U_{\Omega_j},V_{\Omega'_j}$ are the gates on $\vert \Omega_{l_0} \vert$ sites that constitute the approximation. A unitary can be applied to regions $\Omega^*_{\pm l_0}$ to have the circuit act independently on the region of length $2l_0$, and such that the resulting unitary on that region is an approximation to $e^{-i t H_A^{l_0}}$ which, when acting on $A$, gives a good approximation to $e^{-itH}Ae^{itH}$.}
    \label{fig:1DTime}
\end{figure}

The result of \cite{Osborne_2006} is that the time evolution $e^{-itH}$ on $N$ particles can be approximated with a matrix product unitary $M_N^{t}$ of bond dimension $D_t$ such that
\begin{equation}\label{eq:epsilontapprox}
    \vert \vert e^{-itH}-M_N^t \vert \vert \le  \epsilon,
\end{equation}
and 
\begin{equation}
    D_t \le e^{\mathcal{O}(\vert t \vert) + \mathcal{O}\left(\log(N/\epsilon)\right) }.
\end{equation}
The operator $M_N^t$ is in fact a quantum circuit of depth $2$, where the gates act on $ \vert \Omega_N \vert   \equiv \mathcal{O}(\vert t \vert) + \mathcal{O}\left(\log(N/\epsilon)\right)$ adjacent qubits. That is, let $\{\Omega_j\}$ be a partition of the chain into sets of size $\vert\Omega_N\vert$, and let $\{\Omega'_j\}$ be the same partition, displaced by an amount $\vert\Omega_N\vert/2$. Then
\begin{equation}
    M^t_N = \left( \bigotimes_{j=1}^{\left\lceil{N/\vert\Omega_N\vert}\right\rceil} U_{\Omega_j}(t)\right) \left( \bigotimes_{j=1}^{\left\lceil{N/\vert\Omega_N\vert}\right\rceil} V_{\Omega'_j}(t)\right).
\end{equation}
Our local approximation is $M^t_N(l_0)$, which is defined in the same way as $M_N^t$ except for the fact that the partition is into much smaller sets of adjacent sites, each of which has length $\vert \Omega_{l_0}\vert  =\mathcal{O}(\vert t \vert) + \mathcal{O}\left(\log(l_0/\epsilon)\right)$ instead of $\vert \Omega_{N}\vert $. Here $l_0$ is a free parameter such that $\vert \Omega_{l_0}\vert<l_0<N$, otherwise independent of $N$. See Fig. \ref{fig:1DTime} for an illustration of this and the other definitions in the proof.

The goal is to bound the norm $\vert \vert e^{-itH}Ae^{itH}-M^t_N(l_0) A (M^t_N(l_0))^\dagger \vert \vert$, for any operator $A$ with support on at most $k$  adjacent sites. First, with the triangle inequality and the Lieb-Robinson bound,

\begin{align}
    \vert \vert e^{-itH}Ae^{itH}-M^t_N(l_0) A (M^t_N(l_0))^\dagger \vert \vert \le c e^{v_{\text{LR}}t-l}+   \vert \vert e^{-itH^{(l)}_A}Ae^{itH^{(l)}_A}-M^t_N(l_0) A (M^t_N(l_0))^\dagger \vert \vert ,
\end{align}
where $H^{(l)}_A$ is the Hamiltonian containing the terms of $H$ that are a distance $l$ away from $A$, and such that $l_0-k-\vert \Omega_{l_0}\vert \le l <l_0-k$. 

Now, let $\Omega^*_{\pm l_0}$ be the nearest sets at a distance strictly greater than $l$ from $A$, from the left and the right. The key is to notice that, since $M_{l_0}^t$ is a depth-$2$ circuit, there exists a unitary $U_{l_0}\equiv U_{-l_0} \otimes U_{+l_0}$ acting on $\Omega^*_{\pm l_0}$ such that $U_{l_0} M^t_N(l_0)= U \otimes M^t_{l_0}$ for some $U$ with support at distance strictly larger than $l_0$ from A (see Fig. \ref{fig:1DTime}). By construction, the supports of $U_{l_0}$ and $H_A^{l}$ do not overlap.

The MPO $M^t_{l_0}$ is the approximation of the unitary $e^{-i t H^{(l_0)}_A}$, on a region of size $\propto l_0$, which allows us to use Eq. \ref{eq:epsilontapprox}. That is
\begin{align}
    \vert \vert e^{-itH^{(l)}_A}Ae^{itH^{(l)}_A}&-M^t_N(l_0) A (M^t_N(l_0))^\dagger \vert \vert \\ &=   \vert \vert U_{l_0}e^{-itH^{(l)}_A}Ae^{itH^{(l)}_A}U_{l_0}^\dagger-U_{l_0}M^t_N(l_0) A (M^t_N(l_0))^\dagger U_{l_0}^\dagger \vert \vert
    \\ &= \vert \vert e^{-itH^{(l)}_A}Ae^{itH^{(l)}_A}-(U \otimes M^t_{l_0}) A (U \otimes M^t_{l_0})^\dagger \vert \vert =  \vert \vert e^{-itH^{(l)}_A}Ae^{itH^{(l)}_A}- M^t_{l_0} A ( M^t_{l_0})^\dagger \vert \vert
    \\ &\le 2\vert \vert A \vert \vert ce^{v_{\text{LR}}t-l}+ \vert \vert e^{-itH^{(l_0)}_A}Ae^{itH^{(l_0)}_A}- M^t_{l_0} A ( M^t_{l_0})^\dagger \vert \vert
    \\ & \le 2 \vert \vert A \vert \vert ce^{v_{\text{LR}}t-l}+2\epsilon  \vert \vert A \vert \vert
    \\& \le 2 \vert \vert A \vert \vert c e^{v_{\text{LR}}t+k+\vert\Omega_{l_0}\vert-l_0} +2\epsilon \vert \vert A \vert \vert.
\end{align}
In the second line we have used the unitary invariance of the norm, in the third line the fact that $U_{l_0}$ decouples the region at a distance $l$ from the rest, and in the fourth line the Lieb-Robinson bound again, to relate $e^{-itH^{(l)}_A}$ to $e^{-itH^{(l_0)}_A}$.

Thus, since $\vert \Omega_{l_0}\vert \ll l_0 $, if we choose $l_0 = \mathcal{O}\left((k+v_{\text{LR}}t+\log(\frac{1}{\epsilon})\right)$, we achieve the approximation error $ \vert \vert e^{-itH^{(l)}_A}Ae^{itH^{(l)}_A}-M^t_N(l_0) A (M^t_N(l_0))^\dagger \vert \vert\le 3 \epsilon \vert \vert A \vert \vert$. Finally, since $M^t_N(l_0)\equiv M^{t}_{k}$ is by definition a depth-$2$ circuit with gates acting on $\vert \Omega_{l_0} \vert$ sites, the bond dimension required to represent it exactly is 
\begin{equation}
    D_t \le e^{\mathcal{O}\left( \vert t \vert\right)+
    \mathcal{O}\left(\log\left(\frac{k+v_{\text{LR}}t+\log\frac{1}{\epsilon}}{\epsilon}\right) \right)}=e^{\mathcal{O}\left( \vert t \vert\right)}\times \text{poly} \left(\frac{k+v_{\text{LR}}t+\log\frac{1}{\epsilon}}{\epsilon} \right).
\end{equation}

%%%%%%%%%%%%%%%%%%%
%%%%%%%%%%%%%%%%%%

\subsection{Higher dimensions}\label{app:highdtime}

For local approximations in higher dimensions, we cannot straightforwardly generalize the proof of the 1d case because we cannot approximate the unitary evolution with a depth-$2$ circuit, since in higher dimensions one requires depth at least $d+1$ (see Appendix B of \cite{Haah_2021}).

Instead, we use an argument similar to that for thermal states in Appendix \ref{app:localthermal}. We use the global approximation of $e^{-itH}$ described in Appendix \ref{app:clusterexp} to approximate the evolution of the Hamiltonian within the hypercube that has the local operator in the middle. Since this will only work for individual regions, we have to condition the particular partition used on the support of the input, which as we show can be done with a tensor map as defined in Sec. \ref{sec:setting}. 

Let us start with the approximation within each of the hypercubes. Lemma \ref{le:Liebrobinson} guarantees that to simulate the evolution of a local observable $e^{-itH}Ae^{itH}$ where $\text{supp}(A) \subset \mathcal{R}$ we only need the Hamiltonian of a smaller region $H_X$ such that $l:=\text{dist}(A,X^c)\geq 2d-1$. 

We now define the same set of hypercubes covering the lattice as in Appendix \ref{app:localthermal} (also see Fig. \ref{fig:2DThermal})

\begin{equation}\label{eq:hyper2}
    \mathcal{G}_p = \{ \Lambda^{(i)}_{l_0,p}: \cup_i  \Lambda^{(i)}_{l_0,p} = \Lambda \}
\end{equation}
such that  $\vert \Lambda^{(i)}_{l_0,p} \vert \le l_0^d$. Again the parameter $p$ defines e.g one of the corners of the hypercubes, or any other parameter that determines the whole partition. It can adopt $l_0^d$ values. To simulate $A(t)$ we then define the PEPO
\begin{equation}\label{eq:Ml0}
    M_{p}^{(l_0)}=
    \bigotimes_i  M_{i,p}^{(l_0)},
\end{equation}
where $M_{i,p}^{(l_0)}$ is an $\epsilon$ approximation in operator norm to $e^{-itH_i}$, and $H_i =\sum_{x: \mathrm{supp}(h_x)\subseteq \Lambda^{(i)}_{l_0,p}} h_x$ contains the terms of $H$ with support in $\Lambda^{(i)}_{l_0,p}$. Given Eq. \eqref{eq:BDtimeev}, this can be achieved with bond dimension

\begin{equation} \label{eq:BDtimeevlocal}
    D_p \le e ^{\mathcal{O}\left(\vert t \vert d \log{\frac{\vert t\vert d l_0^d}{\epsilon}}\right)}.
\end{equation}
For later convenience we also define the $p=0$ as a special case,
\begin{equation}
M_0^{(l_0)}=\mathbb{I}.
\end{equation}

To approximate $A(t)$ we shall choose the partition $\mathcal{G}_p$ so that there exists a hypercube $\Lambda^{(i_p)}_{l_0,p}$ such that $\text{dist}(A,\setminus \Lambda^{(i_p)}_{l_0,p}) \ge l= l_0/2 -k$ (that is, the region $\mathcal{R}$ is in the middle of the hypercube). This yields an error $
\left(2\epsilon+c  l^{d-1}\e^{v t - l}\right) \norm{A}$, from which we have to then choose $l_0 =\tilde{\mathcal{O}}
\left(k+v_{\text{LR}}t+\log{\frac{1}{\epsilon}} \right)$ to get $3 \epsilon \norm{A}$. 

However, we want a PEPO that approximates \emph{any} A, no matter the region in which it lies.  This can be achieved by constructing a linear map $\mathcal{M}_k^t(A)$ made of tensor contractions such that when applied to $A$ it results on the operator $M_{i,p}^{(l_0)}$ (with the correct partition $p$ and position $i$) being applied as $M_{i,p}^{(l_0)}A (M_{i,p}^{(l_0)})^\dagger$. 

For it to be a good approximation of the dynamics, it should be such that the value of $p$ is \emph{conditioned} on the position of $A$, so that $A$ is in the middle of one of the hypercubes $\Lambda^{(i)}_{l_0,p}$. Thus, it should contain the description of the different $M_{p}^{(l_0)}$ (there are $l_0^d$, one for each value of $p$). Importantly, notice that we need to apply a single $M_{i,p}^{(l_0)}$ and not the full $M_p^{(l_0)}$. Applying $M_p^{(l_0)}$ in every region far from $A$ will incur in additional errors growing with system size, since they are not exactly unitaries and $M_{i,p}^{(l_0)}M_{i,p}^{(l_0) \dagger}\neq \mathbb{I}$. Thus, on top of being able to implement the right partition $p$, we have to make sure that the support beyond the lightcone is trivial.

As we now show, all these requirements can be achieved with a bond dimension $ D^2_p 
\times (l_0+k)^{3d}$, and so we obtain 
\begin{equation}\label{eq:BDhighDT}
    D \le  \tilde{\mathcal{O}}\left(d \vert t \vert\frac{k+v_{\text{LR}}t+\log{1/\epsilon}}{\epsilon}  \right)^{\mathcal{O}(\vert t \vert d )},
\end{equation}
that is, exponential in $t$ (as expected) and polynomial in $k, \epsilon^{-1}$ now with a degree growing with $t$.

We now explain how to construct this PEPO map $\mathcal{M}_k^t(A)$. Let us first limit ourselves to the cases where $A$ is a product of Pauli matrices $A=\bigotimes_x \sigma_x^{(j)}$ with $j \in \{0,1,2,3\}$, which stand for, respectively, the Pauli $\{ \mathbb{I},X,Y,Z\}$ (the generalization to higher local dimensions is straightforward). By definition the only non-identity elements in $A$ have support on a small connected region $\mathcal{R}$ of size $k$. The result extends to arbitrary local operators by linearity.

\textbf{One dimension:} 
We first consider the one-dimensional case, which serves as an illustrative example, and then extend it to larger dimensions. If we simplify the picture by putting the two physical indices of the output and the input together, the tensor network can be drawn schematically as follows.

\begin{figure}[H]
\centering
\includegraphics[width=0.45\linewidth]{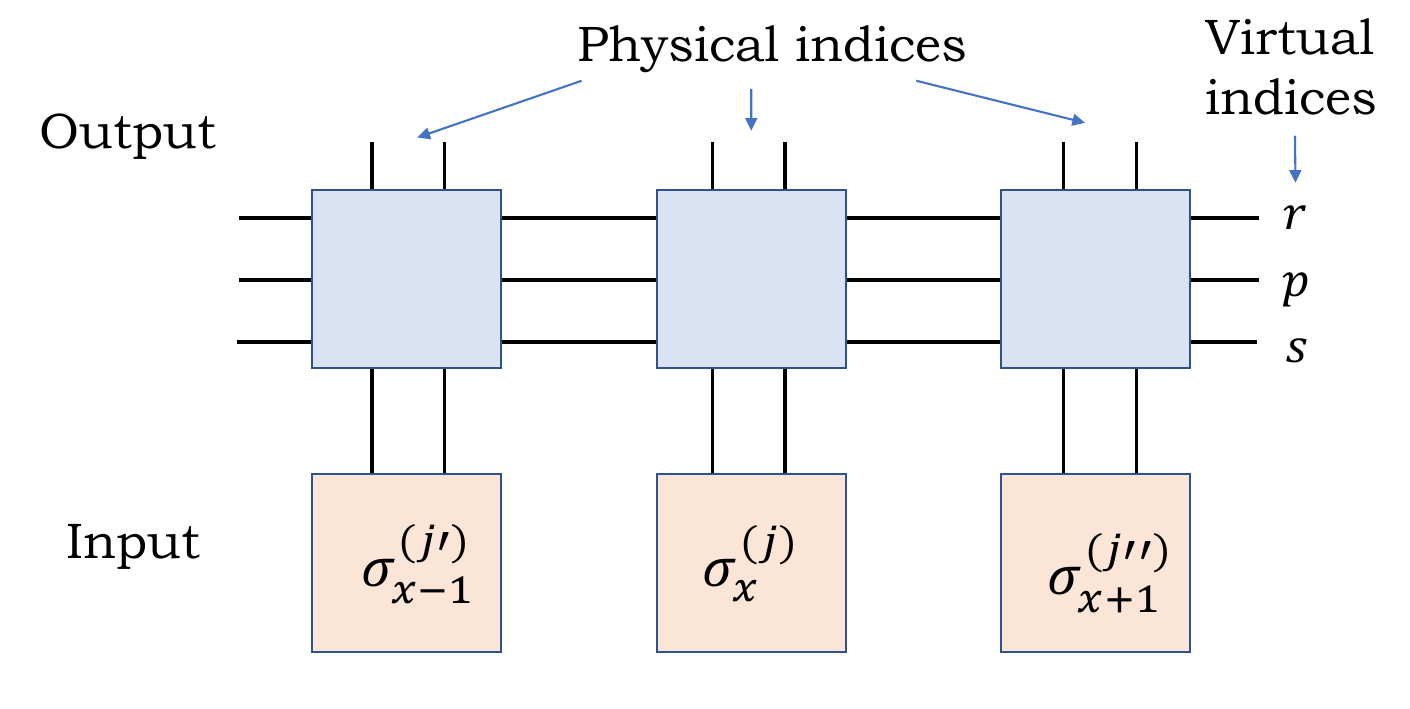}
\label{fig:tensorT4}
\end{figure}
\noindent

We have three virtual indices with different functions: $p$ is the index that determines the partition into hypercubes from Eq. \eqref{eq:hyper2} to use, and as such is fixed by the position of the non-identity Pauli matrices in $A$ (i.e. the region $\mathcal{R}$). It can adopt $l_0+1$ different values. For a given $p$, index $r$ contains the information about the tensors from the MPO $M_p^{(l_0)}$ that are applied to the Pauli, and thus has bond dimension given by Eq. \eqref{eq:BDtimeevlocal}.  Finally, $s=\{s_1,s_2\}$ are the indices that determine the support of the output MPO, ensuring that it is trivial outside the ligthcone, as required. 

The $r$ index is thus determined by the MPOs $M_{l_0}^{(l_0)}$. The $p,s$ indices work as follows: starting from the left of the chain, $s_1$ is zero until it reaches a non-trivial Pauli input at site $x_\mathcal{R}$. At that point it turns into $(l_0+k)/2$, and then decreases by $1$ at each tensor until it reaches a value we label as $0'$, which indicates the end of the lightcone at the right. Beyond that point, the desired outcome $\mathcal{M}_k^t(A)$ must have trivial support, which we enforce by setting the output tensor to be the identity $M_0^{(l_0)}=\mathbb{I}$. The point $x_\mathcal{R}$  also determines the particular value of $p=p^*$, which then sets the PEPO $M_{p^*}^{(l_0)}$ to act within the lightcone. The index $p^*$ is chosen in such a way that the boundary of the hypercubes is $(l_0-k)/2$ away from the left and $(l_0+k)/2$ from the right of $x_\mathcal{R}$. This sets the right part of the lightcone. For the left side of the lightcone, we use index $s_2$. It starts with value $(l_0-k)/2$ at $x_\mathcal{R}$, and decreases by $1$ to the left, while being $0$ all the way to the right. The values from $(l_0-k)/2$ to $1$ to the left of $x_\mathcal{R}$ allow us to set the output tensor to be that of $M_{p^*}^{(l_0)}A (M_{p^*}^{(l_0)})^\dagger$. Then, it goes from $1$ to another $0'$, after which the output of the tensor is the identy, thus setting the end of the lightcone to the right. 

\begin{figure}[H]
    \centering
    \subfigure[]{\includegraphics[width=0.25\textwidth]{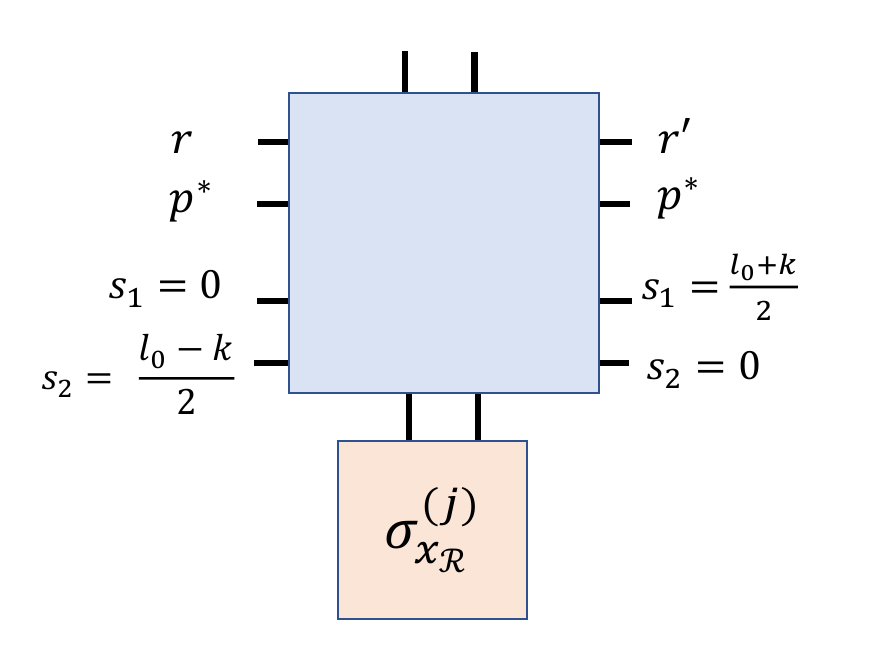}} 
    \subfigure[]{\includegraphics[width=0.25\textwidth]{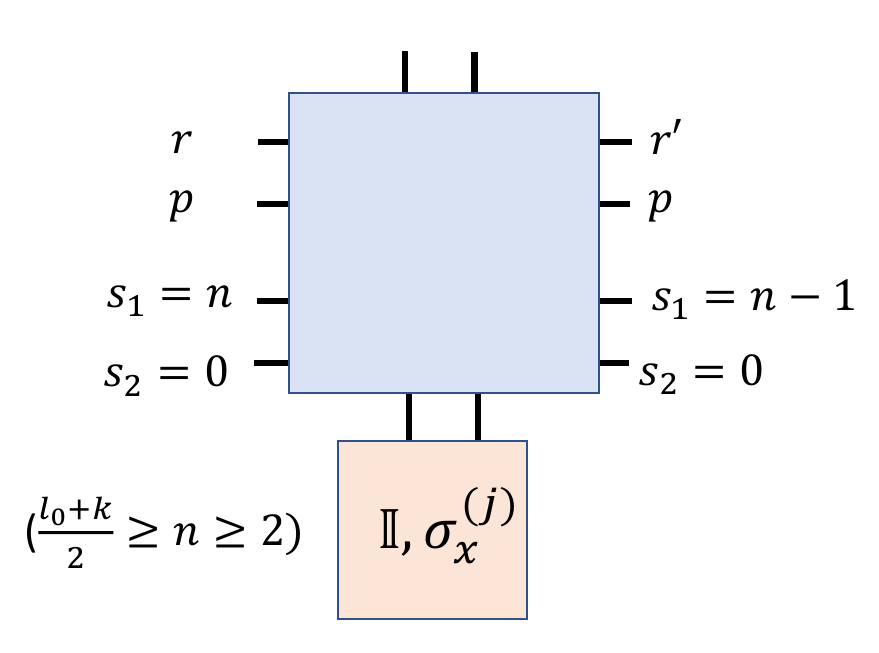}} 
    \subfigure[]{\includegraphics[width=0.25\textwidth]{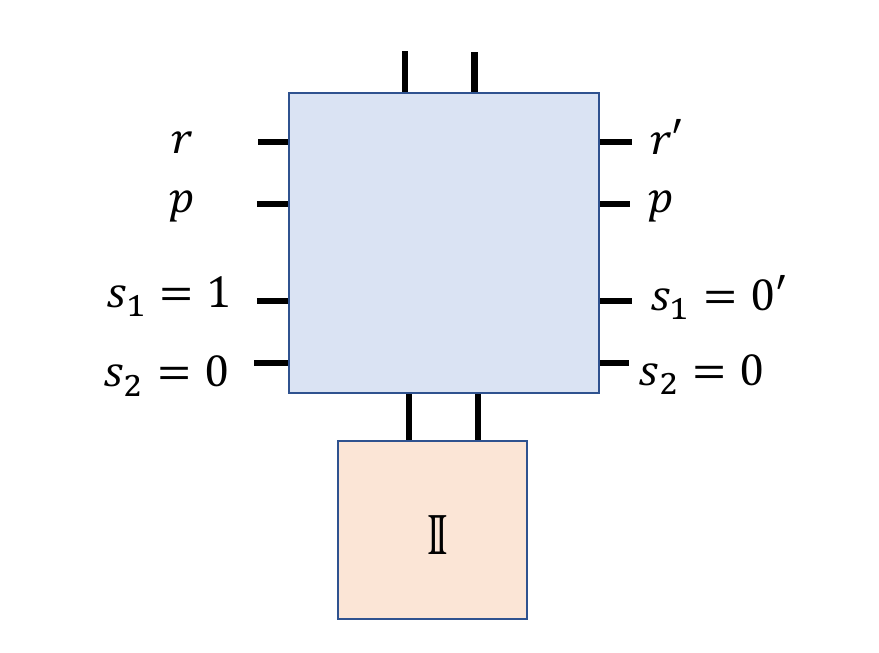}}
    \subfigure[]{\includegraphics[width=0.25\textwidth]{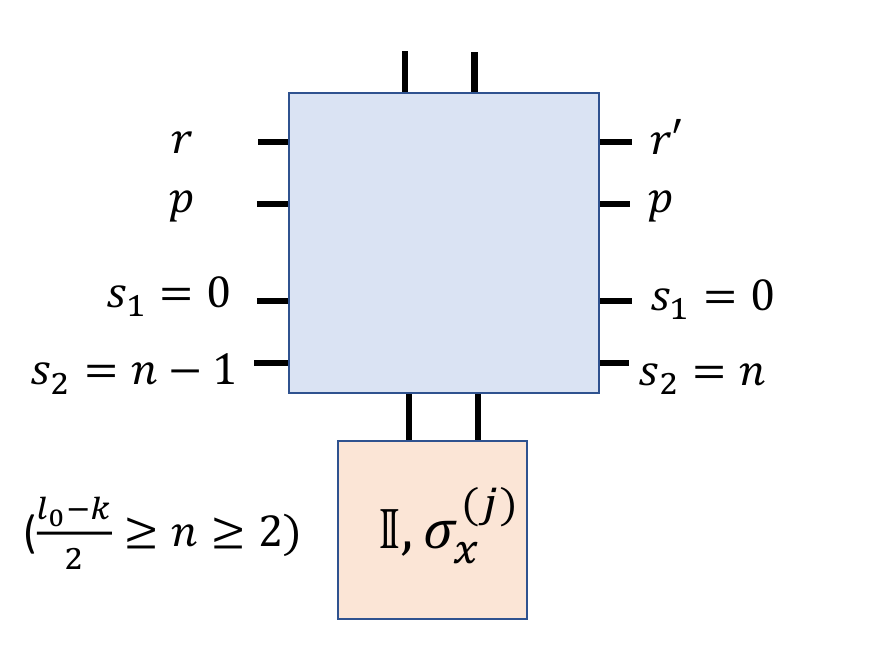}}
    \subfigure[]{\includegraphics[width=0.25\textwidth]{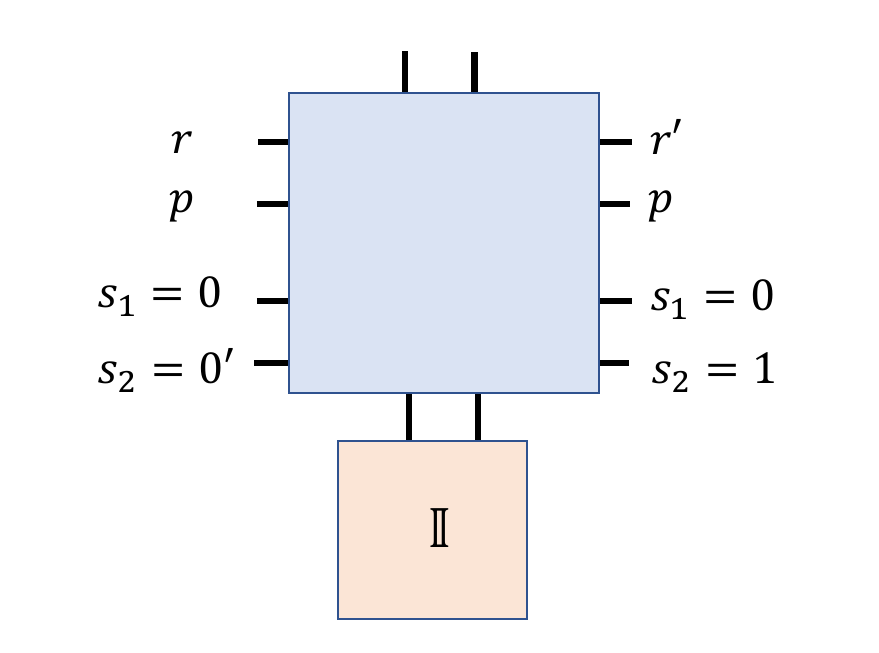}}
    \caption{Tensors: (a)  at position $x_{\mathcal{R}}$ indicating the start of region $\mathcal{R}$, fixing $p=p^*$ (b) to the right of $x_{\mathcal{R}}$ (c) indicating the right end of the lightcone (d) to the left of $x_{\mathcal{R}}$ (e) indicating left end of the lightcone. The $s$ index has been split into $\{s_1,s_2\}$.}
    \label{fig:tensorindv}
\end{figure}

\begin{figure}[H]
\centering
\includegraphics[width=0.8\linewidth]{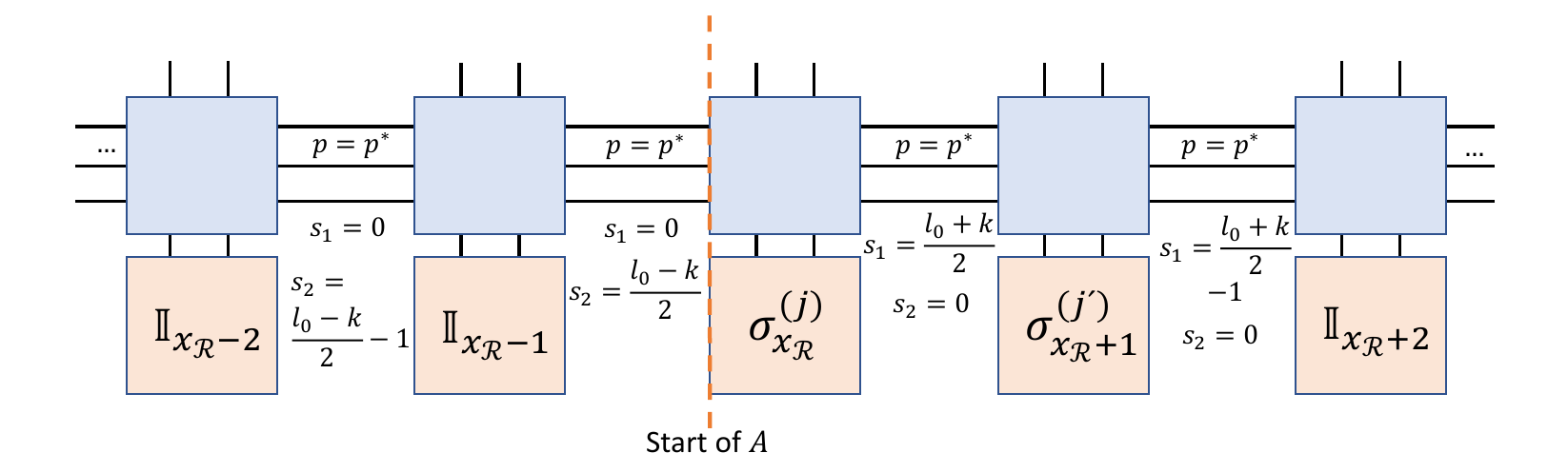}
\caption{Configuration of tensors around $x_{\mathcal{R}}$.}
\label{fig:tensorTin}
\end{figure}

We now illustrate this with the particular tensors. The following combinations of indices are the ones that add to the final result. First, the ones that contribute to the lightcone region are shown in Fig. \ref{fig:tensorindv} as follows.
All of these have as output the tensors corresponding to the MPO $M_{i,p^*}^{(l_0)}A (M_{i,p^*}^{(l_0)})^\dagger$ within the lightcone. The configuration around $x_{\mathcal{R}}$ can be illustrated in Fig. \ref{fig:tensorTin} as follows.

\begin{figure}[H]
    \centering
    \subfigure[]{\includegraphics[width=0.25\textwidth]{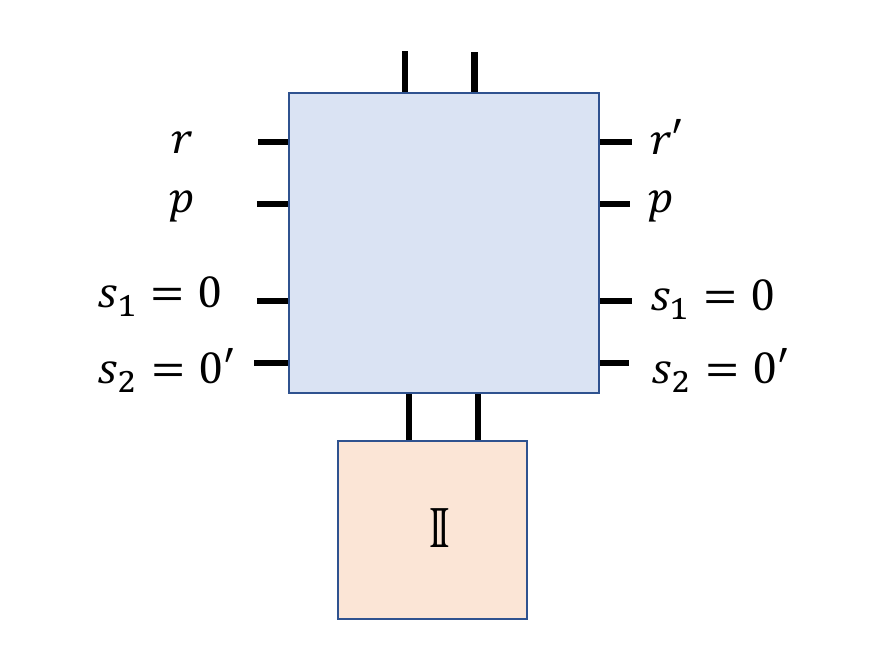}} 
    \subfigure[]{\includegraphics[width=0.25\textwidth]{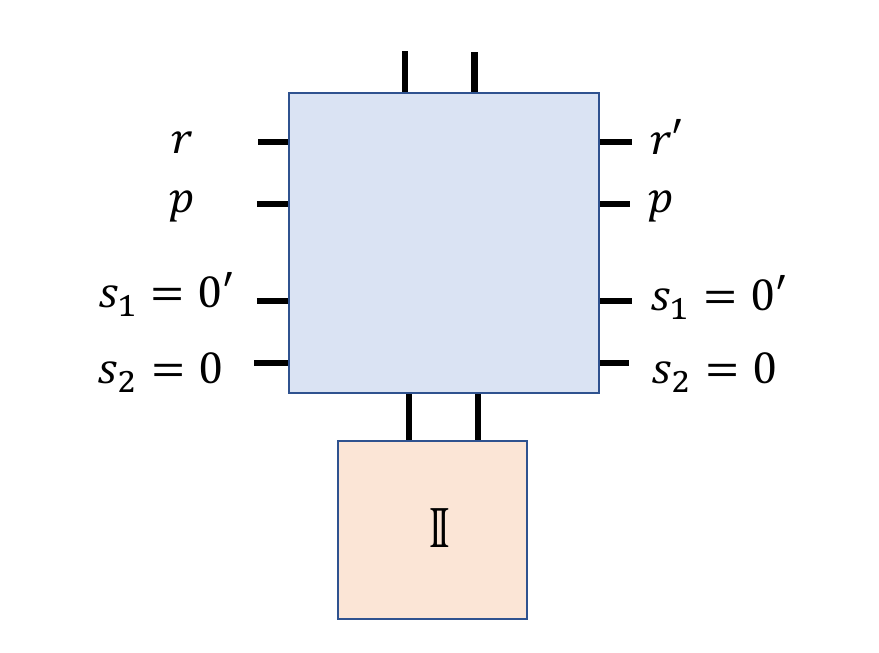}} 
    \caption{Tensors: (a)  to the left and (b) to the right of the lightcone. Their output is the identity.}
    \label{fig:tensorindv2}
\end{figure}

\begin{figure}[b]
    \centering
\includegraphics[width=1.05\textwidth]{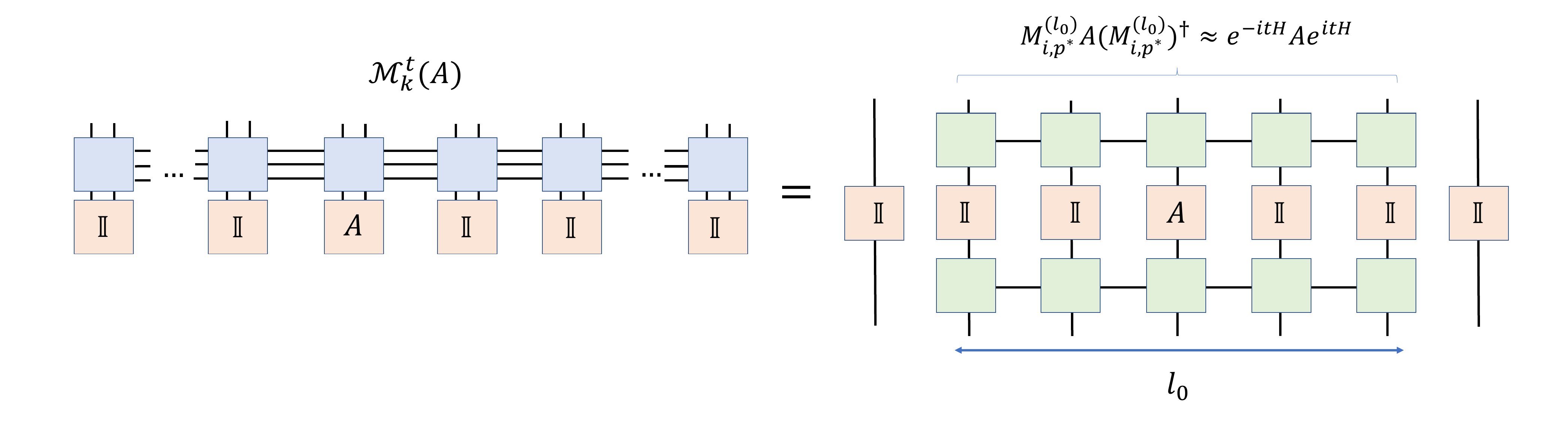} 
    \caption{Final illustration of the scheme in one dimension.}
    \label{fig:tensorfinal}
\end{figure}

\noindent Then, the ones required to set the output beyond the lightcone to be the identity are shown in Fig. \ref{fig:tensorindv2} as the following.
To summarize, $p^*$ determines the right set of hypercubes, and $s_1,s_2$ indicates how far we are to starting point of $\mathcal{R}$, from the right and the left. The tensors apply $M_{p^*}^{(l_0)}$ to $A$ (so that the output is $M_{p^*}^{(l_0)}A (M_{p^*}^{(l_0)})^\dagger$) unless either $s_1,s_2=0'$, in which case they output $\mathbb{I}$. The result is then $M_{i,p^*}^{(l_0)}A (M_{i,p^*}^{(l_0)})^\dagger$ as desired. This is illustrated in Fig. \ref{fig:tensorfinal}.

\textbf{Higher dimensions:} This can be done by establishing the same index configurations as for 1d for every individual direction. Now, we have two $s$ indices $s^{(q)}_1,s^{(q)}_2$ for each dimension $q \in \{X,Y,...\}$, as well as one index $p^{(q)}$ for each. Let us begin with a particular spatial direction, say $q=X$, with coordinates adopting values $x\in \{0,1,...,L_X\}$ ($L_X$ being the length in that direction). Starting from the first position $x=0$, the index $s^X_1$ is $0$ until any non-trivial Pauli appears at the input, with an $X$ coordinate $x_\mathcal{R}$. It is possible that there is more than one non-trivial input with the same initial coordinate $x_\mathcal{R}$, but this does not affect the scheme.

\begin{figure}[H]
\centering
\includegraphics[width=0.6\linewidth]{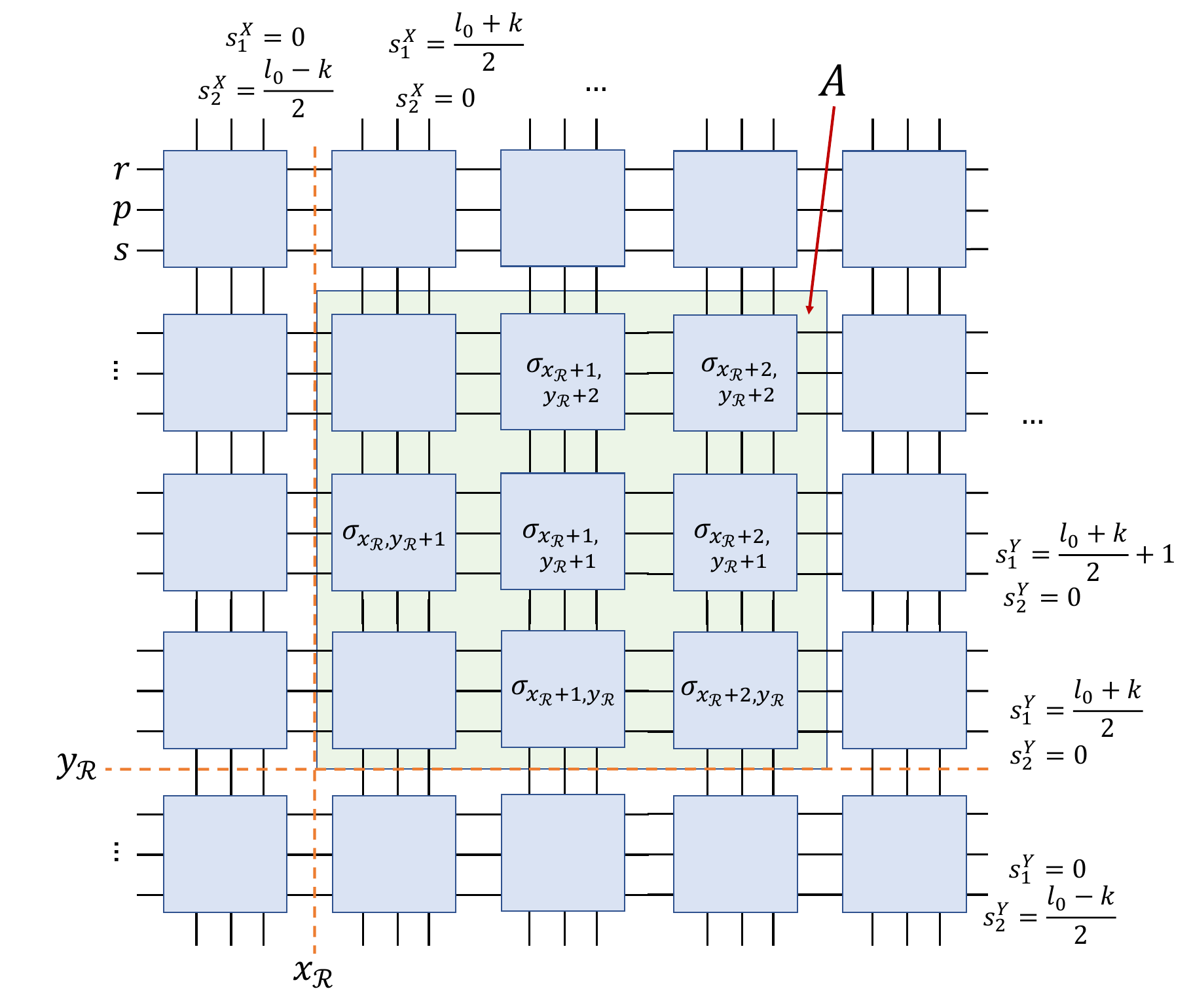}
\caption{In the figure, the $\sigma$ represent sites at which the input is not the identity but any other Pauli matrix. The green shaded region is the square region $\mathcal{R}$ in which the observable $A$ has support. The index $s_1^{X}$ grows to the right, $s_2^X$ decreases to the left, and $s_1^Y,s_2^Y$ respectively grow and decrease downwards. $p^*$ has two components of $l_0+1$ possible values each and is fixed across the entire lattice. We only show virtual indices, and omit the input and output legs of the tensors explicitly for simplicity.}
\label{fig:tensorTin2}
\end{figure}

At $x_{\mathcal{R}}$, the index $s^X_1$ is then again changed to $(l_0+k)/2$, and $s^X_2$ is changed to $(l_0-k)/2$. Beyond $x_{\mathcal{R}}$, $s^X_1$ decreases along the positive $X$ direction, and $s^X_2$ decreases along the negative $X$ direction (while being $0$ in the positive direction). When either $s^X_1=0',s^X_2=0'$, the tensor acts as the identity $M_0^{(l_0)}=
\mathbb{I}$. Also, the position of that first non-trivial Pauli(s) determines the position of the hypercubes in the $X$ direction, as $p^X=p^{X*}$ in the same way as in one dimension. We repeat the same scheme for every dimension. This is illustrated in 2d in Fig. \ref{fig:tensorTin2} as follows.

\begin{figure}[H]
\centering
\includegraphics[width=0.6\linewidth]{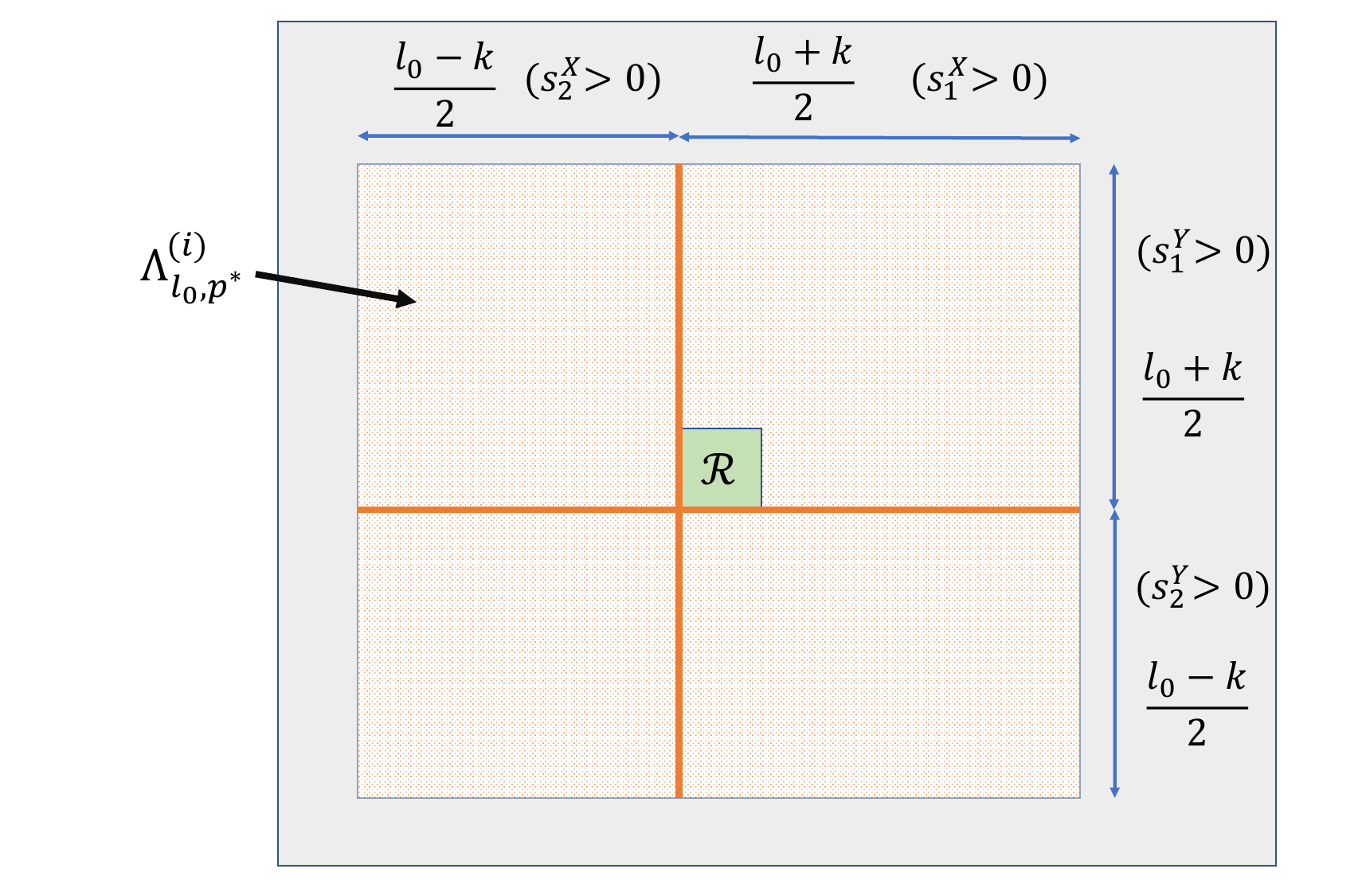}
\caption{The orange region represents the hypercube $\Lambda_{l_0,p^*}^{(i)}$ within the larger lattice in which the output $\mathcal{M}_k^t(A)=M_{i,p^*}^{(l_0)}AM_{i,p^*}^{(l_0)\dagger}$ has support. It is determined by the positivity of indices $s_{1,2}^{(q)}$. When any $s_{1,2}^{(q)}=0'$, we are outside of the hypercube.}
\label{fig:tensorTin3}
\end{figure}

\noindent With this, we obtain a fixed partition into hypercubes as determined by $p^*=\{p^{q*}\}$, and also a set of indices $s$ whose non-zero value indicate where the output light-cone lies. That is, whenever any one of the $2 \times d$  indices $s$ is $0'$, we set the local tensor to be that of $M_0^{(l_0)}$, and otherwise to be $M_{p^*}^{(l_0)}$. This is illustrated in Fig. \ref{fig:tensorTin3} as follows.

Given this scheme, the map outputs $M_{i,p^*}^{(l_0)}AM_{i,p^*}^{(l_0)\dagger}$ with the right $\{i,p^*\}$ chosen, such that $\mathcal{R}$ is in the middle of the corresponding hypercube. From our discussion above, this is an $\epsilon$-close approximation to $e^{-itH}Ae^{itH}$ in operator norm. The whole map involves, on each side of the input $A$, a virtual index $r$ with dimension  $D_p$, and two additional virtual indices $p,\textbf{s}$ with dimension $(l_0+1)^{d}, ((l_0+k)/2)^{2d} $ respectively, to connect the tensors at each site. Thus, the bond dimension can be taken as $D_p^2 \times (l_0+k)^{3d}$, as stated above.

Finally, our result assumed that $A$ was a product of Paulis. Since the map is linear and any local operator can be written as a linear sum of up to $4^{k}$ Pauli matrices, the result follows by replacing $\epsilon \rightarrow 4^{-k} \epsilon$ in the bound for the bond dimension Eq. \eqref{eq:BDhighDT}.

%%%%%%%%%%%%%
%%%%%%%%%%%%%

%%%%%%%%%%%%%%%%%%%%%%%
%%%%%%%%%%%%%%%%%%%%%%

\section{Auto-correlation functions}\label{app:autocorr}

We now study how to approximate the correlation functions in one dimension described in the main text
\begin{equation} \label{eq:autocorrelation2}
    \langle A(t) A \rangle_\beta = \tr{\frac{e^{-\beta H}}{Z} e^{-itH}Ae^{itH} A },
\end{equation}
with the MPOs from our approximations, as 
$\tr{\tilde{\rho}_{k'} M_{k''}^t A M_{k''}^{t \dagger} A} $. Again, $A=\frac{1}{N}\sum_x A_x$, with $A_x$ being supported on $k$ sites and $\norm{A_x} \le \norm{A}$. The region sizes $k'$ is to be determined, and we choose $k''=k$ (the size of the support of $A_x$). 
Throughout the proof, there are different sources of $\epsilon$-size errors (coming from Results \ref{th:thermal} and \ref{th:1Dtime}, and repeated applications of the Lieb-Robinson and the decay of correlations), which we set to be equal.

First, notice that the MPO $M_k^t$ is a depth-2 quantum circuit with the gates acting on $L= \mathcal{O}(\vert t \vert )+\mathcal{O}\left(\log \frac{k+v_{\text{LR}}t+\log(\epsilon^{-1})}{\epsilon} \right)$ sites. Thus, $ M_{k}^t A_x M_{k}^{t \dagger}$ has support on at most $2L$ sites.  Also, by construction,  the MPO $\tilde{\rho}_{k'}$ has the following clustering of correlations property, for arbitrary operators $F,G$ of support smaller than $k'$:
\begin{align}\label{eq:decaycorr1}
    &\vert \tr{\tilde{\rho}_{k'} F G } - \tr{\tilde{\rho}_{k'} F}\tr{\tilde{\rho}_{k'} G }\vert =0 \,\,\, \text{if} \,\,\, \text{dist}(F,G) > \mathcal{O}\left(  \frac{k'}{\epsilon}\right),
    \\ \label{eq:decaycorr2}
    &\vert \tr{\tilde{\rho}_{k'} F G } - \tr{\tilde{\rho}_{k'} F}\tr{\tilde{\rho}_{k'} G } \vert \le(2\epsilon + Ke^{-\text{dist}(F,G)/\xi} )\norm{F}\norm{G} \,\,\, \text{if} \,\,\, \text{dist}(F,G) \le \mathcal{O}\left(  \frac{k'}{\epsilon}\right),
\end{align}
where $\epsilon$ here comes from the error in the approximation in Result \ref{th:thermal}. The first line is due to the fact that $\tilde{\rho}_{k'}$ is constructed as a mixture of partitions, which are product over distances larger than $\mathcal{O}\left(  k'/\epsilon\right)$. The second is due to the fact that, when the observables are close (and within the same cell in the partition), one recovers the correlation length $\xi$ in the thermal state from Eq. \eqref{eq:corrlength}, with an additional error $2\epsilon$.

Let us look now at the pairs $A_x,A_y$ that are far away in the lattice. The previous equations imply  that
\begin{align} \label{eq:errordecay}
 \vert   \tr{\tilde{\rho}_{k'}  M_{k}^t A_x M_{k}^{t \dagger} A_y} - \tr{\tilde{\rho}_{k'}  M_{k}^t A_x M_{k}^{t \dagger}} \tr{\tilde{\rho}_{k'} A_y} \vert \le 3\epsilon \norm{A}^2 \quad \text{if} \quad  \text{dist}(x,y) > \mathcal{O}\left(  \frac{k'}{\epsilon}\right)+\mathcal{O}(v_{\text{LR}}t+k+ \log{1/\epsilon})
\end{align}
which follows from Eq. \eqref{eq:decaycorr1}. Also, from Eq. \eqref{eq:decaycorr2}
\begin{align} \label{eq:errordecay2}
 &\vert   \tr{\tilde{\rho}_{k'}  M_{k}^t A_x M_{k}^{t \dagger} A_y} - \tr{\tilde{\rho}_{k'}  M_{k}^t A_x M_{k}^{t \dagger}} \tr{\tilde{\rho}_{k'} A_y} \vert \le 3\epsilon \norm{A}^2  & \quad \text{if} \quad  \text{dist}(x,y) > \mathcal{O}\left( \xi \log{\frac{1}{\epsilon}}\right)+\mathcal{O}(v_{\text{LR}}t+k+ \log{1/\epsilon}).
\end{align}
Thus, for a good approximation to these terms it suffices to constrain $k' \ge 2L$, in which case 
\begin{align} \label{eq:errordecay3}
 &\vert   \tr{\tilde{\rho}_{k'}  M_{k}^t A_x M_{k}^{t \dagger} A_y} - \langle  M_{k}^t A_x M_{k}^{t \dagger}\rangle_\beta \langle A_y \rangle_\beta \vert \le 7\epsilon \norm{A}^2  & \quad \text{if} \quad  \text{dist}(x,y) > \mathcal{O}\left( \xi \log{\frac{1}{\epsilon}}\right)+\mathcal{O}(v_{\text{LR}}t+k+ \log{1/\epsilon}),
\end{align}
where the $7$ comes from repeated applications of Result \ref{th:thermal}. This thus includes all the $x,y$ pairs with  
\begin{equation}\label{eq:distance} \text{dist}(x,y) > \mathcal{O}\left( \min \left\{ \xi \log{\frac{1}{\epsilon}}, \frac{k'}{\epsilon} \right\}\right) +\mathcal{O}(v_{\text{LR}}t+k+ \log{1/\epsilon}). \end{equation}

Using Result \ref{th:1Dtime}, we have that $\vert \langle  M_{k}^t A_x M_{k}^{t \dagger}\rangle_\beta \langle A_y \rangle_\beta-\langle  A_x(t) \rangle_\beta \langle A_y \rangle_\beta \vert \le 3 \epsilon \vert \vert A \vert \vert^2.$ Repeated applications of the Lieb-Robinson bound yield $ \vert \langle  A_x(t) \rangle_\beta \langle A_y \rangle_\beta - \langle  A_x(t)  A_y \rangle_\beta \vert \le 2 \epsilon \vert \vert A \vert \vert^2$, and so we obtain
\begin{equation} \label{eq:finalcorr}
   \vert  \tr{\tilde{\rho}_{k'}  M_{k}^t A_x M_{k}^{t \dagger} A_y}- \langle  A_x(t)  A_y \rangle_\beta \vert \le 12 \epsilon \vert \vert A \vert \vert^2.
\end{equation}

For the pairs that are nearby, let us now constrain further $k' \ge  \xi + 2L$.  This means that, given Eq. \eqref{eq:distance} we still need to cover the pairs $x,y$ such that 
\begin{equation} \text{dist}(x,y) \le \mathcal{O}(v_{\text{LR}}t+k+ \xi \log{1/\epsilon}). \end{equation}
This can clearly be done by choosing $k' =\mathcal{O}((v_{\text{LR}}t+k+ \xi) \log{1/\epsilon})$, which is consistent with $k' \ge \xi+2L$. Then, Result \ref{th:thermal} implies that 
\begin{equation}
     \left \vert   \tr{\tilde{\rho}_{k'}  M_{k}^t A_x M_{k}^{t \dagger} A_y} - \langle  M_{k}^t A_x M_{k}^{t \dagger} A_y \rangle_\beta \right \vert \le \epsilon \vert \vert A \vert \vert^2,
\end{equation}
and from Result \ref{th:1Dtime}, $ \vert \langle  M_{k}^t A_x M_{k}^{t \dagger} A_y \rangle_\beta - \langle  e^{-itH} A_x e^{itH} A_y \rangle_\beta   \vert \le 3 \epsilon \vert \vert A \vert \vert^2$. This thus shows that Eq. \eqref{eq:finalcorr} holds for all pairs $x,y$ regardless of their distance. The final result follows by approximating each term in the sum $ \langle e^{-itH} A e^{itH}A \rangle_\beta =\frac{1}{N^2}\sum{x,y} \langle e^{-itH} A_x e^{itH}A_y \rangle_\beta$.

%%%%%%%%%%%%%%%%%%%%%%%
%%%%%%%%%%%%%%%%%%%%%%

\end{document}